\documentclass[a4paper,11pt]{article}
\pdfoutput=1 

\usepackage{jheppub} 
\usepackage{amsmath, amsfonts}

\usepackage[T1]{fontenc} 
\usepackage[dvipsnames]{xcolor}
\usepackage{subfigure}
\usepackage{stmaryrd}
\usepackage{enumerate}
\usepackage{tikz}

\usetikzlibrary{calc,intersections,positioning,arrows}

\newcommand{\beq}{\begin{equation}}
\newcommand{\eeq}{\end{equation}}
\newcommand{\bea}{\begin{eqnarray}}
\newcommand{\eea}{\end{eqnarray}}

\DeclareMathOperator{\Tr}{Tr}

\usepackage{color}

\usepackage[normalem]{ulem}  

\newcommand{\non}{\nonumber\\}

\newcommand{\be}{\begin{equation}}
\newcommand{\ee}{\end{equation}}

\title{Softening holographic nuclear matter
}

\author[a]{Christian Ecker,}
\author[b]{Nicolas Kovensky,}
\author[c]{Orestis Papadopoulos,}
\author[c]{Andreas Schmitt}

\affiliation[a]{Institute for Theoretical Physics, Goethe University Frankfurt, 60438 Frankfurt am Main, Germany.}
\affiliation[b]{Instituto de F\'isica de La Plata - CONICET
Diagonal 113 e/ 63 y 64, 1900 - La Plata, Argentina.}
\affiliation[c]{Mathematical Sciences and STAG Research Centre, University of Southampton, Highfield Campus, Southampton
SO17 1BJ, United Kingdom.}

\emailAdd{ecker@itp.uni-frankfurt.de}
\emailAdd{nicolas.kovensky@iflp.unlp.edu.ar}
\emailAdd{op1g19@soton.ac.uk}
\emailAdd{a.schmitt@soton.ac.uk}

\abstract{Baryons in the holographic Witten-Sakai-Sugimoto model are described by instanton solutions on the flavor branes. A commonly used approximation for dense baryonic matter replaces the many-instanton solution by a simpler, spatially  homogeneous, ansatz, which requires a discontinuity in the holographic direction of the non-abelian gauge field in order to account for topological baryon number. We point out that the simplest configuration with a single jump -- often used in previous studies --  results in matter at saturation density that is much stiffer than real-world nuclear matter.
This is improved, although not completely remedied, by adding a second jump. We present a 
systematic discussion of all possible configurations up to four jumps, dynamically computing locations of and behavior at the discontinuities. 
We find solutions that continuously connect to those based on pointlike baryons, thus, for the first time, establishing a concrete link between the instantonic and homogeneous pictures. 
This is supported by translating the multi-jump profiles of the gauge field into gauge invariant multi-layer charge distributions. The most important of our 
novel configurations has a block-like structure in the bulk, becomes pointlike at low density and/or large coupling, and is energetically preferred over all previously studied configurations. 
Therefore, our work lays the ground for improved predictions from holography for dense nuclear matter in neutron stars.}

\begin{document}
\maketitle
\flushbottom

\vspace{-0.5cm}
\section{Introduction}
\label{sec:intro}

Matter at large baryon densities, for instance in the interior of neutron stars, is notoriously difficult to describe theoretically. Effective theories of weakly coupled nuclear matter \cite{Hebeler:2009iv,Drischler:2016djf,Tews:2018kmu} are valid at comparably low densities and can at best be extrapolated to ultra-dense matter. Conversely, perturbative first-principle calculations from Quantum Chromodynamics (QCD) \cite{Freedman:1976ub,Pisarski:1999tv,Kurkela:2009gj} are valid at asymptotically large densities where the QCD coupling constant is small \cite{Gross:1973id,Politzer:1973fx}, and one relies  on extrapolations {\it down} in density. Together with results from numerous phenomenological models and interpolations agnostic to the microscopic physics \cite{Annala:2017llu,Altiparmak:2022bke}, these approaches yield valuable insight into the condensed matter physics of QCD, but are far from providing a complete picture.

A complementary approach is the Anti-de Sitter/Conformal Field Theory (AdS/CFT) correspondence (or ``gauge-gravity duality'') \cite{Maldacena:1997re,Witten:1998qj}, which allows for a non-perturbative calculation valid in the strong-coupling limit. One goal of this approach is to understand the differences between weakly coupled and strongly coupled matter, which can be stark, depending on the observable under consideration. Since a gravity dual of QCD is currently not known and, in most applications, the holographic approach relies on the limit of large number of colors $N_c$, the resulting predictions have to be taken with care. Nevertheless, many previous studies have shown that even on the quantitative level, sometimes surprisingly, 
the gauge-gravity duality can capture real-world non-perturbative physics correctly.
In this paper, we address the holographic description of confined matter at zero temperature and large baryon densities, i.e., ``cold and dense holographic nuclear matter'' within a particular realization of the AdS/CFT correspondence, the  Witten-Sakai-Sugimoto (WSS) model \cite{Witten:1998zw,Sakai:2004cn,Sakai:2005yt}.

This model is a top-down non-supersymmetric construction based on type-IIA string theory, which  incorporates the concept of baryons in AdS/CFT \cite{Witten:1998xy,Gross:1998gk} as D4-branes with $N_c$ string endpoints wrapped on an internal 4-sphere. 
Such objects can be described  equivalently in terms of  gauge field configurations with nontrivial topological charge in the gauge theory on the $N_f$ D8- and $\overline{\rm D8}$-branes \cite{Hata:2007mb,Seki:2008mu}. Here, $N_f$ is the number of flavors, in this paper $N_f=2$.
The preferred single-instanton configuration has its profile centred at the tip of these connected ``flavor branes'', while its width goes to zero as the 't Hooft coupling becomes large. Here the flat-space instanton solution \cite{1975PhLB...59...85B} provides a good approximation, and is the basis for many studies of single baryons in the WSS model, see for instance refs.\  \cite{Hata:2007mb,Seki:2008mu,Hong:2007dq,Hata:2007tn,Hashimoto:2009st}.

Many-baryon systems require, in principle, many-instanton solutions in curved space. This is very difficult to handle, both  analytically and numerically, and thus various degrees of simplifications have been employed, for instance to assume the instantons to be pointlike objects \cite{Bergman:2007wp}. A crystalline structure in the spatial directions is expected for  nuclear matter at large $N_c$ \cite{Klebanov:1985qi,Rho:2009ym,Kaplunovsky:2012gb,Kaplunovsky:2013iza}, and, additionally, the multi-instanton configurations can be expected to spread out in the holographic direction, possibly creating multiple ``layers'' as the density is increased \cite{Kaplunovsky:2012gb,Preis:2016fsp,Elliot-Ripley:2016uwb}. At high densities, the instantons start to overlap, hence assuming a homogeneous configuration in the spatial directions becomes a good approximation. 
Real-world nuclear matter {\it is} homogeneous (a liquid, not a solid), at least around saturation density, which is experimentally accessible in large nuclei \cite{Reinhard:2005nj,Horowitz:2020evx}. Therefore, the homogeneous approximation can also be viewed as a (somewhat phenomenological) 
model approach, which might be closer to actual nuclear matter than the crystalline solutions that are plagued by large-$N_c$ artifacts. 

In particular, one can ask whether holographic matter constructed in this way is able to reproduce basic properties of nuclear matter at saturation (e.g., saturation density, binding energy, and  incompressibility), although the model might strictly speaking predict an inhomogeneous configuration at these relatively low densities. 
It is this homogeneous approach, pioneered in ref.\ \cite{Rozali:2007rx} and pursued in several subsequent works \cite{Li:2015uea,Bartolini:2022gdf,CruzRojas:2023ugm,Bartolini:2025sag}, that we will consider in this paper. This approach is also employed in the so-called V-QCD model \cite{Ishii:2019gta,CruzRojas:2023ugm}, which is a bottom-up realization of the gauge-gravity duality where the ratio $N_c/N_f$ is kept finite \cite{Jarvinen:2011qe,Alho:2012mh,Gursoy:2017wzz}.  We will restrict ourselves to isospin-symmetric nuclear matter, taking a step back from the recent generalization to nonzero isospin \cite{Kovensky:2021ddl,Kovensky:2021kzl,Kovensky:2023mye}, in favor of the following improvements. 

Assuming the gauge fields to be spatially homogeneous does not allow for nontrivial topological charge unless there is at least one discontinuity in the non-abelian sector, along the holographic direction \cite{Rozali:2007rx}. 
The main point of our paper is to determine number and locations of such discontinuities dynamically, considering configurations with up to four jumps.
This includes minimizing the thermodynamic potential with respect to all parameters involved in the behavior of the non-abelian gauge field at each jump. Performing a systematic study allows us to recover some configurations that were already known \cite{Rozali:2007rx,CruzRojas:2023ugm} and also produces many others that have never been discussed before. In particular, some of the energetically favored solutions within the homogeneous ansatz were missed in previous literature. 

Classifying configurations with regard to the number of jumps in the gauge field is a useful tool, but physically of little relevance. We therefore also discuss the emerging gauge invariant charge distributions, which show a layered structure reminiscent of the instantonic approach. As we shall discuss, the number of jumps is {\it not} necessarily identical to the number of layers. In particular, we shall show that two of our novel configurations with three and four jumps, respectively, constitute 1-layer and 2-layer phases, which connect smoothly to their pointlike counterparts -- as the density goes to zero and/or the 't Hooft coupling to infinity. These counterparts are extracted from a general calculation for arbitrarily many pointlike layers, which we present together with a new, semi-analytical, result for the continuum limit of infinitely many pointlike layers.

Our paper is organized as follows. We discuss the general derivation for an arbitrary number of discontinuities in the non-abelian gauge field in section \ref{sec:setup}. This includes the stationarity conditions in section \ref{sec:domega} and the resulting possible boundary conditions of the gauge fields at the discontinuities in section \ref{sec:LRSA}. We begin the discussion of our numerical results in section \ref{sec:onejump}, first pointing out the unphysically large incompressibility of the well-known 1-jump phase in section \ref{sec:D0}, then discussing all possible 2-jump solutions in sections \ref{sec:DL} -- \ref{sec:DA} and their dependence on the 't Hooft coupling in section \ref{sec:phasediagram}. The generalization to larger number of discontinuities is discussed in section \ref{sec:multi}, which contains our main results in the form of the two most relevant configurations in sections \ref{sec:D0L} and \ref{sec:DRL}, the relation to pointlike solutions in section \ref{sec:pointlike}, and the comparison of all phases we consider in section \ref{sec:compare}. We discuss and interpret our results in section \ref{sec:discussion} and give a summary and outlook in section \ref{sec:summary}.

\section{Setup}
\label{sec:setup}

\subsection{Action}
\label{sec:action}

We consider the confined geometry of the Witten-Sakai-Sugimoto model with antipodal separation of the flavor branes, and we assume that their embedding in the background geometry is geodesic and fixed\footnote{This is a stable configuration, at least in the absence of gauge fields \cite{Sakai:2004cn}. We do not see an obvious reason why the baryonic solutions studied in this paper should change this. To prove stability of the probe branes rigorously, however, fluctuations on the background of the numerical gauge field solutions developed here must be considered. A full stability investigation (including other potential instabilities of our solutions, e.g., with respect to spatial inhomogeneities) is beyond the scope of this paper.}. Moreover, we use the Yang-Mills (YM) approximation for the SU(2) gauge theory on the flavor branes, and include a Chern-Simons (CS) contribution that accounts for the topological baryon number. We write the gauge field action as     
\be\label{action}
S= {\cal N} N_f \frac{V}{T}\int_{u_{\rm KK}}^\infty du \, {\cal L} \, , 
\ee
where $V$ is the three-volume and $T$ the temperature (which only appears in the prefactor, none of the physics we discuss depends on $T$). The prefactor $V/T$ arises from the (here trivial) space-time integral, assuming that the gauge fields in the Lagrangian are uniform. The remaining nontrivial integral is over the holographic coordinate $u\in [u_{\rm KK},\infty]$, where, in our units,
\be
u_{\rm KK} = \frac{4}{9}
\ee
 arises from the compactification of a ``fifth'' extra dimension with the inverse radius given by the Kaluza-Klein mass $M_{\rm KK}$. The integral in eq.\ (\ref{action}) is over one half of the connected flavor branes. The other half gives the same result and the factor 2 is already taken into account in the prefactor, which includes the integral over the internal four-sphere,  
\be
{\cal N} \equiv \frac{N_c M_{\rm KK}^4\lambda_0^3}{6\pi^2} \, ,
\ee
where 
\be
\lambda_0\equiv \frac{\lambda}{4\pi} \, ,
\ee
with the 't Hooft coupling $\lambda$. Except for ${\cal N}$, $V$, and $T$, which combine to a numerical prefactor, we have written the action in terms of dimensionless quantities. For a derivation from the original action and for more background information about the model in the context of dense matter, including generalizations to the Dirac-Born-Infeld (DBI) action and the deconfined geometry, see for instance ref.\ \cite{Kovensky:2021ddl} and references therein. 

The dimensionless version of the Lagrangian is 
\be\label{Lag}
{\cal L} = \frac{u^{5/2}}{2\sqrt{f}}(g_1+g_2-fA'^2)-\frac{3\lambda_0}{4} A(h^3)' \, , 
\ee
where the first and second terms are the YM and CS contributions, respectively, while 
\be
f(u) = 1-\frac{u_{\rm KK}^3}{u^3}  
\ee
is the blackening factor of the metric, and
\be
g_1\equiv \frac{3fh'^2}{4} \, , \qquad g_2 \equiv \frac{3\lambda_0^2h^4}{4u^3} \, .
\ee
Here, prime denotes derivative with respect to $u$. As can be seen from the general derivation \cite{Kovensky:2023mye}, within our homogeneous ansatz for isospin-symmetric matter it is consistent with the equations of motion to restrict ourselves to two independent nonzero components of the gauge field $A_\mu = \hat{A}_\mu {\bf 1}+ A_\mu^a\sigma_a$ (with $\mu=0,1,2,3,u$ and the Pauli matrices $\sigma_a$, $a=1,2,3$): the temporal abelian component, which, for notational convenience, we denote by 
\be
A(u)\equiv \hat{A}_0(u) \, , 
\ee
and the non-abelian spatial field whose components are given by the function\footnote{Nonzero isospin induces a difference between the components $A_i$ and  a nonzero  $\hat{A}_i$ \cite{Kovensky:2023mye}.} $h(u)$,
\be
A_i^a(u) = -\frac{\lambda_0}{2}h(u)\delta_i^a \, , 
\ee
with $i=1,2,3$. In particular, we work in the gauge where $A_u=0$.

It is convenient to introduce a new holographic coordinate $z\in [-\infty,\infty]$ that covers both halves of the flavor branes and that is related to $u$ via
\be \label{uz}
u = (u_{\rm KK}^3+u_{\rm KK}z^2)^{1/3} \, ,
\ee
such that $z=0$ corresponds to the tip of the connected flavor branes at $u=u_{\rm KK}$. 
We shall use both $z$ and $u$ coordinates depending on which one is more convenient and sometimes even employ both in the same equation for notational compactness. To avoid confusion, however, primes will only be used for derivatives with respect to $u$ and derivatives with respect to $z$ will be written out explicitly. 

In all configurations we consider, we employ the boundary conditions \cite{Rozali:2007rx,Li:2015uea}
\be\label{bcs}
A(z=\pm\infty) = \bar{\mu}_B \, , \qquad h(z=\pm \infty) = 0 \,, 
\ee
where $\bar{\mu}_B$ is the dimensionless chemical potential; its dimensionful counterpart, the physical baryon chemical potential, is 
\be \label{mumu}
\mu_B = N_c\lambda_0M_{\rm KK}\bar{\mu}_B \, .
\ee

\subsection{Baryon number from discontinuities}
\label{sec:nB}

The CS term creates topological baryon number of the form 
\be\label{nBrho}
\bar{n}_B = -\frac{1}{2}\frac{3}{4\lambda_0^2} \int_{-\infty}^\infty dz\, \Tr[F_{ij}F_{kz}]\epsilon_{ijk} = \frac{1}{2}\frac{3\lambda_0}{4} \int_{-\infty}^\infty dz\, \partial_z(h^3)\, , 
\ee
where the following components of the field strength tensor have been used, \be
F_{ij}^a = -\epsilon_{ija}\frac{\lambda_0^2 h^2}{2} \, , \qquad F_{iz}^a = \delta_{ia}\frac{\lambda_0}{2}\partial_z h \, .
\ee
It will be instructive later to discuss the $z$-dependent contribution to the baryon density of the various configurations. To this end, we define the (dimensionless and normalized) charge distribution
\be \label{rhoz} 
\rho(z) \equiv  \frac{3\lambda_0}{8\bar{n}_B}\partial_z(h^3) \, , \qquad \int_{-\infty}^\infty dz\, \rho(z) = 1 \, .
\ee
The dimensionful, physical baryon number is given by\footnote{The dimensionless baryon number $\bar{n}_B$ is half of the total baryon density, indicated by the explicit factor $1/2$ in eq.\ (\ref{nBrho}). This is notationally convenient for the calculation, which is performed on one half of the flavor branes. The conversion factor to the dimensionful baryon density in eq.\ (\ref{nBnB}) ensures that $n_B$ is the physical baryon density with contributions from both halves. } 
\be\label{nBnB}
n_B = \frac{\lambda_0^2M_{\rm KK}^3 }{3\pi^2} \bar{n}_B \, .
\ee
We see that $\bar{n}_B$ is given by the integral over a total derivative. Hence, due to the boundary condition in eq.\ (\ref{bcs}), $\bar{n}_B$ can only be nonzero if $h(z)$ has at least one discontinuity. This discontinuity mimics the topological winding of the more general solution that depends on $z$ and the spatial position $\vec{x}$. We shall see below that the topological expression (\ref{nBrho}) indeed gives the same result for $\bar{n}_B$  as the one obtained from the asymptotic behavior of $A(z)$ at the holographic boundary, as it should be according to the AdS/CFT dictionary, and the same as the one obtained from the  thermodynamic definition via the derivative of the pressure with respect to $\bar{\mu}_B$. 

\begin{figure} [t]
\begin{center}
\includegraphics[width=0.5\textwidth]{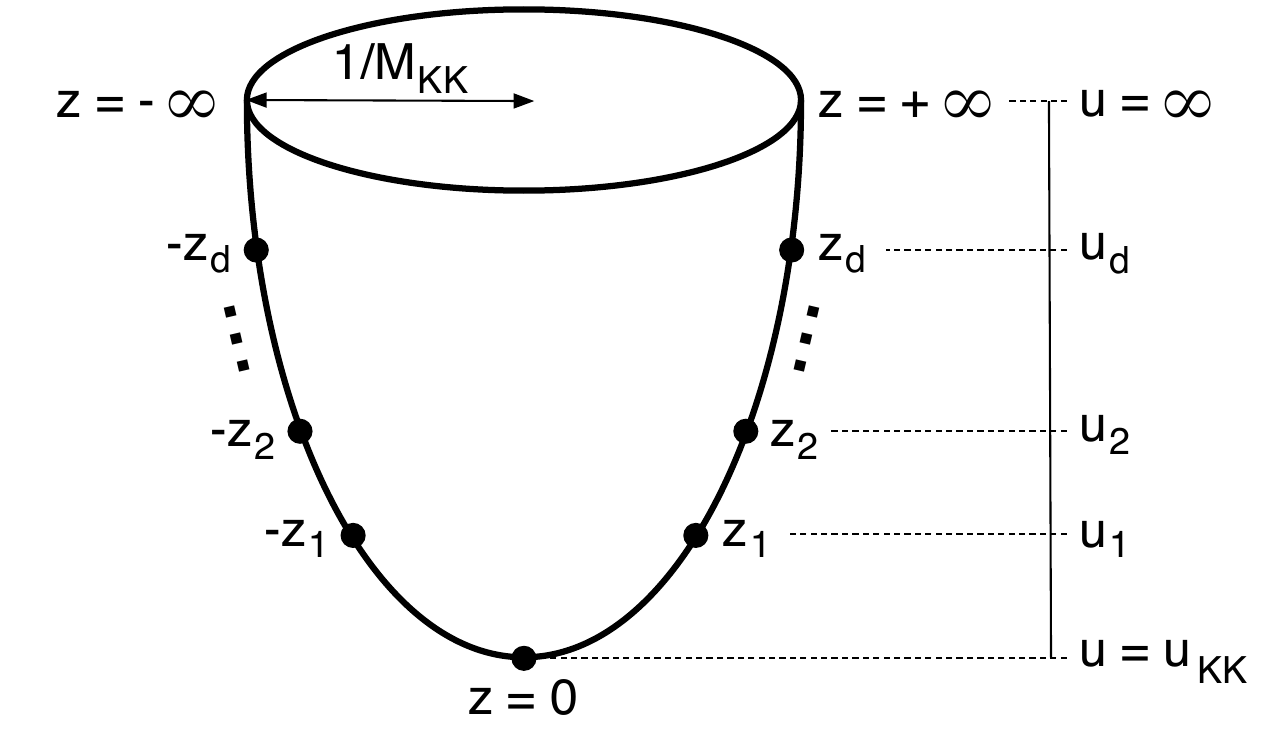}
\caption{Illustration of geometry and notation in both $u$ and $z$ coordinates. We allow for discontinuities of the non-abelian gauge field at the tip $z=0$ of the connected flavor branes and at dynamically determined locations $\pm z_k$. In the general derivation we include arbitrarily many discontinuities, keeping $d$ general, in the concrete numerical results we allow for up to four, i.e., $d\le 2$. For example, a 3-jump configuration has jumps at $z=0$ and $z=\pm z_1$, while a 4-jump configuration has jumps at $\pm z_1$ and $\pm z_2$ (and none at $z=0$).}  
\label{fig:jumps}
\end{center}
\end{figure} 

In general, let us assume that $h(z)$ has $d$ discontinuities on each half of the flavor branes and potentially one additional discontinuity at $z=0$. We label the discontinuities at $z>0$ by $k=1,\ldots, d$, and denote their locations by $z_k$ (correspondingly by $u_k$ in the $u$ coordinate), assuming $z_1< z_2 <\ldots < z_d$, see figure \ref{fig:jumps}.  The values of $h$ at the discontinuities are denoted by 
\be
h_c\equiv h(0^+) \, , \qquad h_{k\pm} \equiv h(z_k^\pm) \, , \quad  k=1,\ldots, d \, .
\ee
If $h(z)$ was even in $z$, the contribution to the baryon density from the discontinuities would cancel pairwise. Hence $h(z)$ must be odd in $z$, i.e., $h(-z_k^\mp) = - h(z_k^\pm)$ and $h(0^-)=-h(0^+) $. Moreover,  $h_c = h(0) = 0$ if there is no jump at $z=0$. With this notation and employing eqs.\ (\ref{nBrho}) and (\ref{rhoz}), the baryon density is 
\be \label{nBsum}
\bar{n}_B = \frac{3\lambda_0}{4}\left[\sum_{k=1}^d (h_{k-}^3-h_{k+}^3)-h_c^3\right] \, .
\ee
In our concrete results we shall study configurations with 1, 2, 3, and 4 jumps, i.e., $d=0$ with nonzero $h_c$, $d=1$ with vanishing and nonvanishing $h_c$, and $d=2$ with $h_c=0$.

\subsection{Equations of motion and free energy}
\label{sec:eom}

The conjugate momenta of the fields $A(u)$ and $h(u)$ are  
\begin{subequations}
\bea
\pi_A &=& \frac{\partial {\cal L}}{\partial A'} = -u^{5/2}\sqrt{f} A' \, , \\[2ex]
\pi_h &=& \frac{\partial {\cal L}}{\partial h'} = \frac{3u^{5/2}\sqrt{f}h'}{4}-\frac{9\lambda_0}{4}Ah^2  \label{Pih}\, , 
\eea
\end{subequations}
and we derive the equations of motion
\begin{subequations} \label{EOM12}
\bea
u^{5/2}\sqrt{f}A' &=& \bar{n}_B Q \, , \label{EOM1} \\[2ex]
(u^{5/2}\sqrt{f}h')' &=& \frac{\lambda_0^2 h^2}{u^{1/2}\sqrt{f}}\left(2h+\frac{3\bar{n}_BQ}{\lambda_0u^2}\right)\, . \label{EOM2}
\eea
\end{subequations}
We have already integrated the equation of motion for $A$, introducing the function $Q(u)$, which is defined as  
\be \label{Q}
\bar{n}_B Q(u) \equiv \frac{3\lambda_0}{4}\int_{u_{\rm KK}}^u dv\, (h^3)'= \frac{3\lambda_0}{4}\left[h^3(u) + \sum_{k=1}^{d} \Theta(u-u_k)(h_{k-}^3-h_{k+}^3)-h_c^3\right] \, , 
\ee
such that $Q(u_{\rm KK})=0$ and $Q(\infty)=1$. The integrated form (\ref{EOM1}) shows that the leading behavior of the derivative of the gauge potential is $A'(u) =  \bar{n}_B/u^{5/2} + \ldots$, confirming that the baryon density is given by the asymptotic behavior. 

Despite the discontinuities in $h(u)$, the function $Q(u)$ is continuous: First of all, at the tip of the connected branes we switch to the $z$ coordinate and continuity is obvious because $Q(z=0)=0$. Then, picking an arbitrary point where $h(u)$ is discontinuous, say $u_\ell$, we find 
\bea
\bar{n}_B Q(u_\ell^+) &=& \frac{3\lambda_0}{4}\left[h_{\ell +}^3+\sum_{k=1}^{\ell} (h_{k-}^3-h_{k+}^3)-h_c^3\right] \non[2ex]
&=& \frac{3\lambda_0}{4}\left[h_{\ell -}^3+\sum_{k=1}^{\ell-1} (h_{k-}^3-h_{k+}^3)-h_c^3\right] = \bar{n}_B Q(u_\ell^-) \, .
\eea
As a consequence, $\pi_A(u)$ [and thus $A'(u)$] is continuous. We shall also assume $A(u)$ itself to be continuous. One might argue that $A(u)$ can be allowed to have jumps, given that we have introduced discontinuities anyway through $h(u)$. However, we keep in mind that the discontinuities of $h(u)$ are an artifact that is unavoidable within our homogeneous ansatz. We expect the ``full'' solution (instantonic solution including spatial dependence) to be smooth, resulting in a smooth temporal abelian component $A(u)$. Therefore, it makes sense to retain the continuity of $A(u)$ if possible -- and we shall see that it is indeed consistent with our equations of motion to do so. We can thus use the shorthand notation  
\be
A_c\equiv A(u_{\rm KK}) \, , \qquad A_k\equiv A(u_k) \, , \quad  k=1,\ldots, d \, , 
\ee
without having to distinguish between the values $A(u_k^+)$ and $A(u_k^-)$. The continuity of $A(u)$ in fact yields conditions needed later for solving the system. Considering each segment between two discontinuities except for the ``last'' one, $u>u_d$, we have the $d-1$ conditions ($k=2,\ldots, d$)
\begin{align}
\begin{aligned}
\label{Acont}
A_k-A_{k-1} = &\int_{u_{k-1}}^{u_k} du\, A' = \frac{3\lambda_0}{4}\int_{u_{k-1}}^{u_k} du\, \frac{h^3}{u^{5/2}\sqrt{f}} \\
&+\frac{\lambda_0}{2u_{\rm KK}^{3/2}}\left[\sum_{\ell=1}^{k-1}(h_{\ell-}^3-h_{\ell+}^3)-h_c^3\right]\left(\arctan\frac{z_k}{u_{\rm KK}} - \arctan\frac{z_{k-1}}{u_{\rm KK}}\right)  \, . 
\end{aligned}
\end{align}
If there is also a discontinuity at $u=u_{\rm KK}$, we have the additional condition
\begin{equation}
\label{A0Ac}
A_1-A_c =  \frac{3\lambda_0}{4}\int_{u_{\rm KK}}^{u_1} du\, \frac{h^3}{u^{5/2}\sqrt{f}} -\frac{\lambda_0h_c^3}{2u_{\rm KK}^{3/2}}\arctan\frac{z_1}{u_{\rm KK}} \, .
\end{equation}  
The segment $u>u_d$ is different because it contains the chemical potential as a boundary condition and  will thus be treated separately.  
For the configurations we consider later, these conditions are only relevant for 3 and 4 jumps, because for 1 or 2 jumps there is no segment between $u_1$ (or $u_{\rm KK}$) and $u_d$.

The equations of motion (\ref{EOM12}) decouple, hence the nontrivial  numerical calculation reduces to solving the differential equation (\ref{EOM2}) for $h(u)$ (it turns out to be somewhat advantageous to do so in the $z$ variable). The result can then be used to compute the function $A(u)$ via eq.\ (\ref{EOM1}) by a straightforward numerical integration. For our purposes, however, the function $A(u)$ is of minor relevance.  What we do need to compute is the chemical potential. Due to the boundary condition for $A$ in eq.\ (\ref{bcs}) we can compute $\bar{\mu}_B$ from 
\bea \label{muB}
\bar{\mu}_B &=& A_d + \int_{u_{d}}^\infty du\, \frac{\bar{n}_BQ}{u^{5/2}\sqrt{f}} \non[2ex]
&=& A_d  +\frac{\bar{n}_B}{3u_{\rm KK}^{3/2}}\left(\pi- 2\arctan\frac{z_d}{u_{\rm KK}}\right) + \frac{3\lambda_0}{4}\int_{u_{d}}^\infty du\, \frac{h^3}{u^{5/2}\sqrt{f}} \, , 
\eea
where, in the second step, we have used eqs.\ (\ref{nBsum}) and (\ref{Q}). [If we are interested in a given $\bar{\mu}_B$, we need to solve eq.\ (\ref{muB}) simultaneously with eq.\ (\ref{EOM2}).] 

Once the equations of motion are solved we can compute the free energy 
density from the on-shell action,
\be
N_f{\cal N} \Omega = \frac{T}{V}\left.S\right|_{\rm on-shell}  \, , 
\ee
where we have introduced the dimensionless free energy density 
\be \label{OmL}
\Omega = \int_{u_{\rm KK}}^\infty du \, {\cal L} \, .
\ee
To derive a convenient form of $\Omega$ we first compute 
\be
\frac{3\lambda_0}{4}\int_{u_{\rm KK}}^\infty du \,A(h^3)' = \bar{\mu}_B \bar{n}_B - \int_{u_{\rm KK}}^\infty du\,\frac{(\bar{n}_BQ)^2}{u^{5/2}\sqrt{f}} \, , 
\ee
where we have employed partial integration and used the continuity of $A(u)$, together with eqs.\ (\ref{nBsum}) and (\ref{Q}). By means of eqs.\ (\ref{action}) and (\ref{Lag}), this yields  
\be \label{Omega}
\Omega = \int_{u_{\rm KK}}^\infty du\, \frac{u^{5/2}}{2\sqrt{f}}\left[g_1+g_2+\frac{(\bar{n}_BQ)^2}{u^5}\right]-\bar{\mu}_B \bar{n}_B.
\ee
The integral is finite and does not require any renormalization (since we have started from the YM action on the flavor branes, purely geometric contributions that are the same for all phases we consider are not part of $\Omega$). The vacuum solution $h(u)=0$ corresponds to $\Omega=0$.  

\subsection{Stationarity conditions}
\label{sec:domega}

So far, we do not have sufficient information to solve the equation of motion (\ref{EOM2}) because each segment between two discontinuities requires boundary conditions, and because the locations of these discontinuities are not yet fixed, except when there is only one. Each discontinuity away from $z=0$ is given by three parameters: its location $z_k$ and the two values $h_{k+}$ and $h_{k-}$. Due to the antisymmetry of $h(z)$, the discontinuity at $z=0$ is special in the sense that it is given by the single parameter $h_c$. We will determine all these parameters dynamically, i.e., find the stationary points of the free energy with respect to them. To this end, we compute the derivative of the free energy with respect to any of the variables 
$x\in \{\bar{\mu}_B, h_c, h_{k\pm}, u_k\}$, with all other variables in this set held fixed. We have included the thermodynamic variable $\bar{\mu}_B$ in this set to confirm thermodynamic consistency within the same calculation. Most conveniently, the derivatives are computed by starting from the general form (\ref{OmL}) with ${\cal L} = {\cal L}(A,A',h,h')$. The variables $u_k$ need a separate treatment and will be discussed below. For all other variables we find with the help of the equations of motion 
\bea \label{dOmdx}
\frac{\partial \Omega}{\partial x} &=& \int_{u_{\rm KK}}^\infty du\, \left[\partial_u\left(\pi_A\frac{\partial A}{\partial x}\right)+\partial_u\left(\pi_h\frac{\partial h}{\partial x}\right)\right] \non[2ex]
&=& -\bar{n}_B\frac{\partial \bar{\mu}_B}{\partial x} + \sum_{k=1}^d\left[\pi_h(u_k^-)\frac{\partial h_{k-}}{\partial x} - \pi_h(u_k^+)\frac{\partial h_{k+}}{\partial x}\right] - \pi_h(u_{\rm KK})\frac{\partial h_c}{\partial x}  \, ,
\eea
where, in the second step, we have used $\pi_A(u_{\rm KK})=0$, $\pi_A(\infty) = \bar{n}_B$, $A(\infty)=\bar{\mu}_B$, $h(\infty)=0$, and the fact that $A(u)$ is continuous. The latter ensures that stationarity with respect to $A_c$ and $A_k$ is automatically fulfilled. 
For $x=\bar{\mu}_B$ we can now easily read off the expected thermodynamic  relation
\be
\frac{\partial \Omega}{\partial \bar{\mu}_B} = -\bar{n}_B \, .
\ee
It remains to evaluate the conditions 
\be \label{dOmhh}
\frac{\partial \Omega}{\partial h_c} = \frac{\partial \Omega}{\partial h_{k\pm}} = 0 \, . 
\ee
Stationarity with respect to $h_{k+}$ and $h_{k-}$ gives the two conditions 
\begin{subequations} \label{eqsUpDown}
\bea
u_k^{5/2} \sqrt{f(u_k)} \, d_{k+}&=& 3\lambda_0 h_{k+}^2 A_k \, , \label{eqUp} \\[2ex]
u_k^{5/2} \sqrt{f(u_k)} \, d_{k-}&=& 3\lambda_0 h_{k-}^2 A_k  \,  .   \label{eqDown}
\eea
\end{subequations}
where we have used eq.\ (\ref{Pih}) and abbreviated
\be
d_{k\pm}\equiv h'(u_k^\pm) \, .
\ee
For the derivative with respect to $h_c$ we need to be more careful since $h'(u)$ diverges at $u=u_{\rm KK}$. In the vicinity of this point we have \cite{Kovensky:2021ddl},
\be \label{hexp}
h(u) = h_c + h_{(1)}\sqrt{u-u_{\rm KK}} + \ldots \, , 
\ee
with a constant $h_{(1)}$ that has to be determined later. 
With this behavior we find from the stationarity with respect to $h_c$,
\be\label{h1Ac}
u_{\rm KK}^2h_{(1)} = 2\sqrt{3}\, \lambda_0 h_c^2 A_c \, .
\ee

We emphasize that eq.\ (\ref{dOmhh}) and the resulting conditions (\ref{eqsUpDown}) and (\ref{h1Ac}) are only valid if a discontinuity at the respective point is assumed. If $h(z)$ is continuous at a certain point $z_k$, the value $h(z_k)$ is given unambiguously by the boundary conditions and the equation of motion. Only if the function is ``broken up'' at $z_k$, new disconnected segments of $h(z)$ arise, and additional conditions, namely eq.\ (\ref{dOmhh}), are required. 
Therefore, although a continuous $h(z)$ at $z=0$ with $h_c = 0$ and $h_{(1)}\neq 0$ seemingly violates eq.\ (\ref{h1Ac}), this  does {\it not} indicate an instability towards a nonzero $h_c$.

Finally, we require stationarity with respect to the locations of the discontinuities $u_k$. Since they appear in the boundaries of the piecewise integration, this case gives the more complicated result 
\bea
\frac{\partial \Omega}{\partial u_k} &=& \int_{u_{\rm KK}}^\infty du\, \left[\partial_u\left(\pi_A\frac{\partial A}{\partial u_k}\right)+\partial_u\left(\pi_h\frac{\partial h}{\partial u_k}\right)\right]+\left.{\cal L}\right|_{u=u_k^-} -\left.{\cal L}\right|_{u=u_k^+} \non[2ex] 
&=& \sum_{\ell=1}^d\left[\pi_A(u_\ell^-)\left.\frac{\partial A}{\partial u_k}\right|_{u_\ell^-} - \pi_A(u_\ell^+)\left.\frac{\partial A}{\partial u_k}\right|_{u_\ell^+} +\pi_h(u_\ell^-)\left.\frac{\partial h}{\partial u_k}\right|_{u_\ell^-} - \pi_h(u_\ell^+)\left.\frac{\partial h}{\partial u_k}\right|_{u_\ell^+}\right]\non[2ex]
&&+\left.{\cal L}\right|_{u=u_k^-} -\left.{\cal L}\right|_{u=u_k^+} \, .
\label{dOmdul}
\eea
Now we need to take into account that $h_{\ell\pm}=h(u_\ell^\pm)$ may depend on $u_k$ explicitly and, if $k=\ell$, via the functional dependence on $u$. The latter has to be subtracted to be left with the explicit dependence that is needed for the boundary terms in eq.\ (\ref{dOmdul}), i.e., 
\be
\left.\frac{\partial h}{\partial u_k}\right|_{u_\ell^\pm} = \frac{\partial h_{\ell\pm}}{\partial u_k}-\delta_{\ell k} h'(u_k^\pm) \, . 
\ee
The first term on the right-hand side does not contribute since $h_{\ell\pm}$ are independent variables, with respect to which we have already extremized the free energy. Together with the analogous relation for $A$ instead of $h$ we thus find that the sum in eq.\ (\ref{dOmdul}) collapses to the term where $\ell=k$, and we are left with 
\bea
\frac{\partial \Omega}{\partial u_k} &=& (\pi_A A'+\pi_h h'-{\cal L})_{u=u_k^+} - (\pi_A A'+\pi_h h'-{\cal L})_{u=u_k^-} \, .
\eea
Here we have not used the continuity of $\pi_A$ yet. We rather first simplify our result by observing that, for any $u$,
\be
\pi_A A'+\pi_h h' - {\cal L} = \frac{u^{5/2}}{2\sqrt{f}} \left[\frac{3fh'^2}{4} - \frac{3\lambda_0^2h^4}{4u^3}-\frac{(\bar{n}_BQ)^2}{u^5}\right] \, .
\ee
Consequently, stationarity with respect to $u_k$ gives 
\bea \label{statu0}
u_k^3f(u_k)(d_{k+}^2-d_{k-}^2) = \lambda_0^2(h_{k+}^4-h_{k-}^4) \, .
\eea
In contrast to eqs.\ (\ref{eqsUpDown}) and (\ref{h1Ac}), this equation is trivially fulfilled if there is no jump at $u_k$.

To summarize: For a potential discontinuity at $z=0$ we need to obey  the condition (\ref{h1Ac}), while for each discontinuity at $z>0$ we need to obey (\ref{eqsUpDown}) and (\ref{statu0}). If the number of jumps is greater than 2, these conditions have to be supplemented by eqs.\ (\ref{Acont}) and/or (\ref{A0Ac}). 
Together with the equation of motion (\ref{EOM2}) this yields a closed system that can be solved numerically. This will become more explicit when we discuss specific classes of solutions in sections \ref{sec:onejump} and \ref{sec:multi}.

\subsection{Behavior at the discontinuities}
\label{sec:LRSA}

For each $z>0$ discontinuity there are four classes of solutions, which is seen as follows.   
We can fulfil one of the eqs.\ (\ref{eqUp}) and (\ref{eqDown}) trivially by setting $d_{k+}=h_{k+}=0$ or $d_{k-}=h_{k-}=0$,  i.e., by requiring $h(u)$ to vanish on one side of the jump. For $d_{k+}=h_{k+}=0$, eq.\ (\ref{eqUp}) is trivially fulfilled and eqs.\ (\ref{eqDown}) and (\ref{statu0}) give a condition between value and derivative of $h$ at the nontrivial  side of the jump, as well as an expression for the gauge potential at this point, 
\be \label{Ljump}
\lambda_0 h_{k-}^2 = u_k^{3/2}\sqrt{f(u_k)}\,d_{k-} \, , \qquad A_k = \frac{u_k}{3} \, .
\ee
Here we have assumed that $d_{k-}>0$, otherwise $|d_{k-}|$ would appear in the first relation. We shall assume $d_{k\pm}>0$ throughout the paper (we did not find any numerical solutions with $d_{k\pm}<0$). Since for the case (\ref{Ljump}) $h(u)$ is only nontrivial ``to the left'' of the jump (i.e., for $u<u_k$) we shall refer to this case as the {\bf left ($L$) discontinuity}. For a single jump on each half of the flavor branes, this configuration was already discussed in ref.\ \cite{Rozali:2007rx}. 

Analogously, we obtain the {\bf right ($R$) discontinuity} by setting $ d_{k-} = h_{k-} = 0$, and
\be \label{Rjump}
\lambda_0 h_{k+}^2 = u_k^{3/2}\sqrt{f(u_k)}\,d_{k+} \, , \qquad A_k = \frac{u_k}{3} \, .
\ee
Again, the configuration where there is  a single $R$ jump on each half of the flavor branes was already discussed in the literature \cite{CruzRojas:2023ugm}. 

If both sides of the jump are nontrivial, eqs.\ (\ref{eqsUpDown}) and (\ref{statu0}) yield the condition 
\be \label{twofactors}
(h_{k+}^4-h_{k-}^4)[\lambda_0^2h_{k-}^4-u_k^3f(u_k)d_{k-}^2] = 0 \, .
\ee
This provides two additional classes of solutions, one for each of the factors being zero. In both classes, $h(u)$ is nonzero on both sides of the jump. Requiring the first factor to vanish gives
\be \label{Sjump}
h_{k+}=-h_{k-} \, , \qquad d_{k_+} = d_{k-} \, , \qquad A_k = \frac{u_k^{5/2}\sqrt{f(u_k)}\,d_{k+}}{3\lambda_0 h_{k+}^2} \, .
\ee
We refer to this case as the {\bf symmetric ($S$) discontinuity}. 

\begin{figure} [t]
\begin{center}
\includegraphics[width=\textwidth]{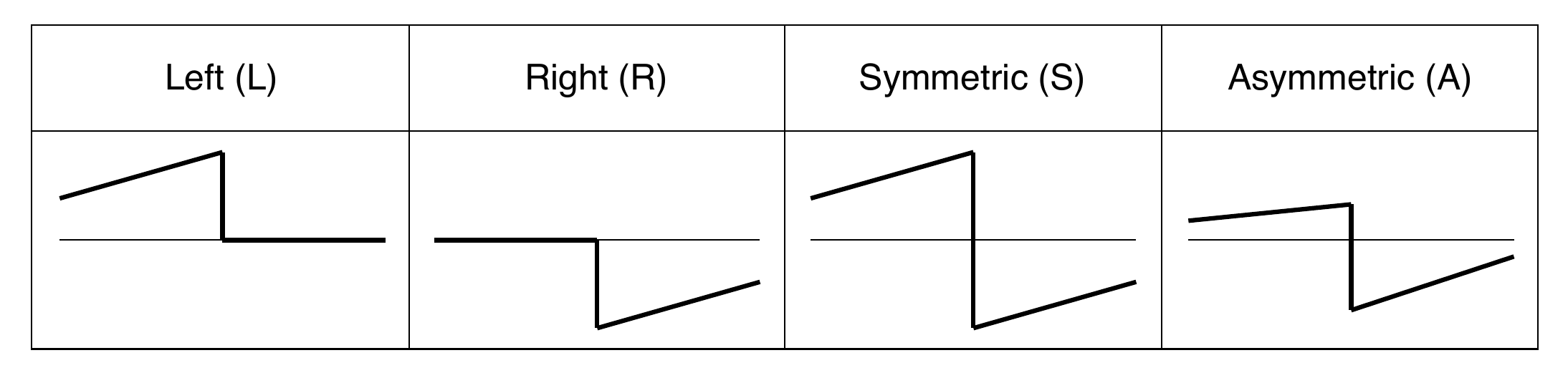}
\caption{Sketch and terminology for the 4 different configurations at a discontinuity that correspond to stationary points of the free energy. The lines illustrate the non-abelian gauge field in the form of the function $h(z)$ with $z>0$ increasing in each panel from left to right. }  
\label{fig:LRSA}
\end{center}
\end{figure}

If the second factor in eq.\ (\ref{twofactors}) vanishes, we obtain a configuration which can be understood as a combination of the $L$ and $R$ jumps, 
\be \label{Ajump}
\lambda_0 h_{k\pm}^2 = u_k^{3/2}\sqrt{f(u_k)}\,d_{k\pm} \, , \qquad A_k = \frac{u_k}{3} \, .
\ee
The value and the derivative of $h(u)$ on each side of the jump are related to each other, but there is no local constraint across the two sides. Therefore, we refer to this configuration as the {\bf asymmetric ($A$) discontinuity}. As a summary, and for further reference, we have collected all four cases in figure \ref{fig:LRSA}. 

These four behaviors, together with a possible discontinuity at $z=0$, are the building blocks for all phases we consider. We shall denote a phase with an even number of discontinuities by $D_{x_1x_2\ldots x_d}$, where $x_k\in \{L,R,S,A\}$, i.e., the shapes of the jumps on one half of the flavor branes are indicated in the subscript, in the order of increasing $z$. For instance, a phase with two jumps (one on each half of the branes) of the $L$ type is denoted by $D_L$. Phases with a discontinuity at $z=0$ have an odd number of jumps and are denoted by $D^0_{x_1x_2\ldots x_d}$. For instance, the phase with three jumps where the $z\neq 0$ jumps are of the $S$ type is denoted by $D^0_S$.

\section{Configurations with one and two jumps}
\label{sec:onejump}

\subsection{$D^0$ phase and incompressibility}
\label{sec:D0}

\begin{figure} [t]
\begin{center}
\hbox{\includegraphics[width=0.5\textwidth]{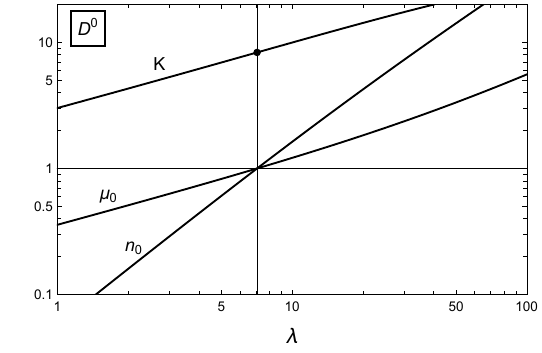}\includegraphics[width=0.5\textwidth]{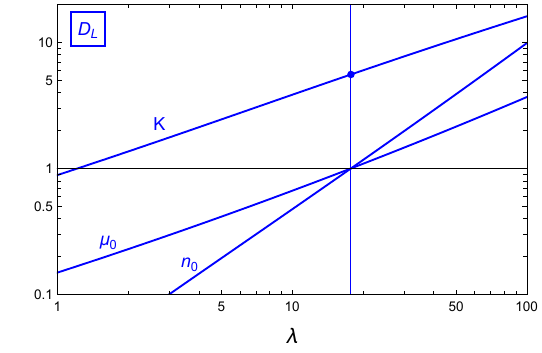}}
\caption{Onset chemical potential $\mu_0$, saturation density $n_0$, and incompressibility $K$ at saturation, as a function of the 't Hooft coupling $\lambda$ for the $D^0$ (left) and $D_L$ (right) phases. Each quantity is normalized by its physical value
(the upper limit $K=300\, {\rm MeV}$ of the empirical range for the incompressibility), such that 1 (thin horizontal line) indicates the physical value. Each panel is created with the value of $M_{\rm KK}$ needed to reproduce the physical values for $n_0$ and $\mu_0$, i.e., $(\lambda,M_{\rm KK}) = (7.09,1000\, {\rm MeV})$ (left) and $(\lambda,M_{\rm KK}) = (17.8,520\, {\rm MeV})$ (right); the corresponding values of $\lambda$ are indicated by a thin vertical line. The dots on these lines mark the incompressibilities that are 8.3 times (left) and 5.6 times (right) larger than the (largest conceivable) physical value at saturation.}  
\label{fig:Komp}
\end{center}
\end{figure}

The phase with a single jump -- $D^0$ in the terminology just introduced -- is the simplest configuration. To compute its properties, we solve eq.\ (\ref{EOM2}) numerically for a given $h_c$ [which corresponds to fixing $\bar{n}_B$ via eq.\ (\ref{nBsum})] and the boundary condition in eq.\ (\ref{bcs}). The resulting function $h(u)$ is then used to compute $\bar{\mu}_B$ from eq.\ (\ref{muB}) (with $A_d$, $z_d$, $u_d$ replaced by $A_c$, 0, $u_c$) and the free energy density (\ref{Omega}). In the dimensionless units of the previous subsections, this calculation requires us to choose a value for $\lambda$, while the second model parameter $M_{\rm KK}$ can be chosen after the numerical calculation to translate the result into physical units. We shall not present a detailed analysis of the $D^0$ phase because this was already done for instance in refs.\
\cite{Li:2015uea,Kovensky:2021ddl,Kovensky:2021kzl}. However, we discuss some properties of this phase to set the stage for our main results. Firstly, we recall that there is a first-order transition  from the vacuum  to the $D^0$ phase (the ``first-order baryon onset''). Furthermore, there exists a choice of the model parameters, namely \cite{Kovensky:2021kzl}
\be \label{lamMKK}
\lambda \simeq 7.09 \, , \qquad M_{\rm KK}\simeq 1000\, {\rm MeV} \, , 
\ee
 for which the holographic calculation reproduces the physical onset chemical potential $\mu_0 \simeq 922.7\, {\rm MeV}$ (the binding energy of nuclear matter at saturation is $E_B\simeq- 16.3\, {\rm MeV}$) and the saturation density $n_0\simeq 0.15\, {\rm fm}^{-3}$. The parameters  (\ref{lamMKK}) are of the same order, but not identical, to the parameters obtained from a fit to the rho meson mass and the pion decay constant in the original works of the WSS model \cite{Sakai:2004cn,Sakai:2005yt}. This tension between fitting to medium properties vs.\ vacuum properties was realized for instance in refs.\ \cite{Kovensky:2021kzl,Kovensky:2021wzu,Kovensky:2023mye}. In two of these works, the $D^0$ phase -- extended to isospin-asymmetric matter -- was used to construct neutron stars.  Although basic properties of the holographic stars such as mass, radius, and tidal deformability were in agreement with empirical data, the stiffness of low-density nuclear matter in the isospin-symmetric $D^0$ phase is much larger than given by experimental data. We demonstrate this discrepancy by computing the incompressibility 
\be \label{Kdef}
K  = 9n_B \frac{\partial \mu_B}{\partial n_B} \, ,
\ee
whose empirical range at saturation density is  \cite{glendenning,Blaizot:1995zz,Vretenar:2003qm,Youngblood:2004fe,Shlomo:2006ole,Liu:2024qds} 
\be
K(n_0)\simeq (200-300)\, {\rm MeV} \, .
\ee
With the parameters (\ref{lamMKK}) we find  $K(n_0)\simeq 2500\, {\rm MeV}$ in the $D^0$ phase. In the left panel of figure \ref{fig:Komp} we show how $K(n_0)$, together with $n_0$ and $\mu_0$, varies with $\lambda$ at fixed $M_{\rm KK}$. Variations of $M_{\rm KK}$ induce  simple rescalings according to $\mu_0, K\propto M_{\rm KK}$, $n_0\propto M_{\rm KK}^3$. The plot shows that nuclear matter can be made softer by decreasing $\lambda$, reaching realistic values for very small values of the coupling. However, this  also decreases $\mu_0$ and $n_0$, and there is no choice of $\lambda$ and $M_{\rm KK}$ where all three basic properties $K$, $\mu_0$, $n_0$ are within, say, 10\% of their real-world values.

Allowing for two jumps in $h(u)$ gives rise to the phases $D_L$, $D_R$, $D_S$, $D_A$. They can be thought of as allowing the location of the $z=0$ discontinuity in the $D^0$ phase to move ``upwards'' in the holographic direction; then, by symmetry, a second discontinuity must appear on the other half of the flavor branes.  Compressibility measures how well the system accommodates particle number without a large increase in pressure $P=-\Omega$. One might thus expect that allowing for an additional freedom, namely allowing the location of the jump(s) to adjust dynamically, would make  it ``easier'' for the system to add baryon number, thereby reducing its incompressibility. Indeed, this is borne out in the $D_L$ phase, as the right panel of figure \ref{fig:Komp} shows. As the $D^0$ phase, the $D_L$ phase shows a first-order baryon onset, and there is a set of parameters that reproduces the physical $n_0$ and $\mu_0$, namely $\lambda=17.8$, $M_{\rm KK} = 520\, {\rm MeV}$. With these values we find that at saturation $K\simeq 1670\, {\rm MeV}$, significantly lower than in the $D^0$ configuration, but still much larger than in nature.  

Let us explain how this result is obtained and, more generally, let us discuss all four phases with two discontinuities and their basic properties. For all two-jump solutions there is a single unknown location $z_1$. Working on the $z>0$ half of the flavor branes we may write 
\be \label{hupdown}
h(z) = \left\{\begin{array}{cl} h_L(z) & \;\;\mbox{for}\;\; 0 < z < z_1\,, \\ h_R(z) & \;\;\mbox{for}\;\; z_1 < z < \infty\,.\end{array}\right.
\ee

\subsection{$D_L$ phase}
\label{sec:DL}

In the $D_L$ phase, $h(z)$ vanishes for all $z>z_1$, i.e., $h_{1+}= h_R(z)=0$, while there is a nonzero value $h_{1-}$ on the ``left'' side of the jump. This value is in one-to-one correspondence with the baryon density. Indeed, from eq.\ (\ref{nBsum}) we obtain 
\be \label{nBh0} 
\bar{n}_B = \frac{3\lambda_0}{4} h_{1-}^3 \, .
\ee
This relation implies  $h_{1-}>0$ since we are only interested in positive baryon number. The initial conditions for $h_L(z)$ are
\be \label{initialL}
h_L(0) = 0 \, , \qquad \left.\frac{\partial h_L}{\partial z}\right|_{z=0} = \frac{h_{(1)}}{\sqrt{3u_{\rm KK}}} \, ,
\ee
where the derivative follows from eq.\ (\ref{hexp}) and the relation between $u$ and $z$, see eq.~(\ref{uz}). (In the $u$ variable one would have to deal with an infinite derivative at $u=u_{\rm KK}$ and thus a regulator in the numerical calculation, which is possible but inconvenient.) The boundary conditions at the discontinuity are 
\be \label{boundaryL} 
h_L(z_1) = h_{1-} \, , \qquad \left.\frac{\partial h_L}{\partial z}\right|_{z=z_1} = \frac{2u_{\rm KK}^{1/2}\lambda_0h_{1-}^2}{3u_1^2} \,, 
\ee
 where the condition for the derivative comes from eq.\ (\ref{Ljump}). We also need, see eq.\ (\ref{Q}), 
\be
\bar{n}_BQ(z) = \frac{3\lambda_0}{4} [h^3(z) + h_{1-}^3\Theta(z-z_1)] \, .
\ee
For a given $h_{1-}$,  we can now solve the equation of motion (\ref{EOM2}) with initial conditions (\ref{initialL}) simultaneously with the conditions (\ref{boundaryL}) for the function $h_L(z)$ and the variables $h_{(1)}$ and $z_1$. Since $h(z)$ vanishes for $z>z_1$, the chemical potential assumes a particularly simple form. With $A_1 = u_1/3$ from eq.\ (\ref{Ljump}), the general form of $\bar{\mu}_B$ (\ref{muB}) leads to 
\be \label{muBDL}
\bar{\mu}_B = \frac{u_1}{3} + \frac{\bar{n}_B}{3u_{\rm KK}^{3/2}}\left(\pi-2\arctan\frac{z_1}{u_{\rm KK}}\right) \, . 
\ee
Inserting $h(z)$ and $\bar{\mu}_B$ into eq.\ (\ref{Omega}) then gives the free energy density.  

\begin{figure} [t]
\begin{center}
\hbox{\includegraphics[width=0.33\textwidth]{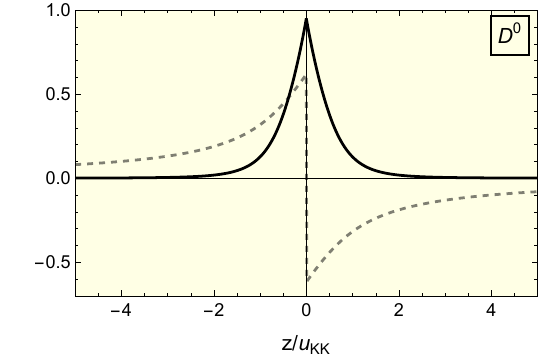}\includegraphics[width=0.33\textwidth]{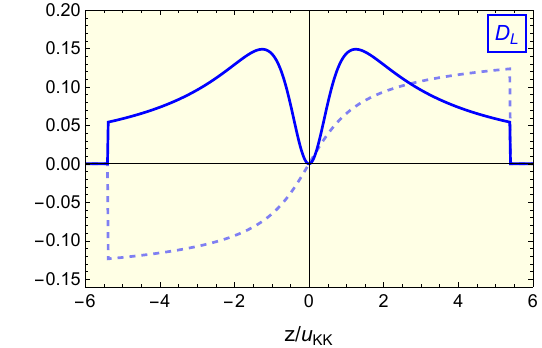}\includegraphics[width=0.33\textwidth]{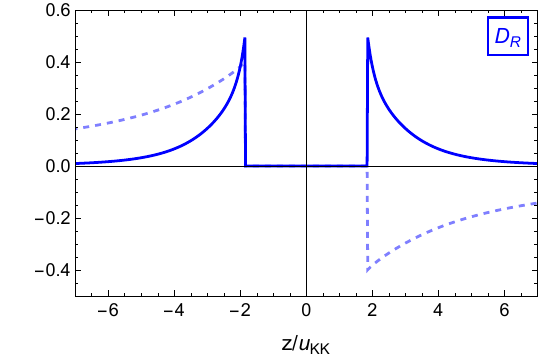}}

\hbox{\includegraphics[width=0.33\textwidth]{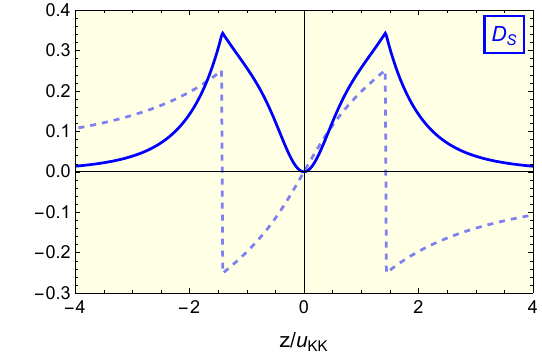}\includegraphics[width=0.33\textwidth]{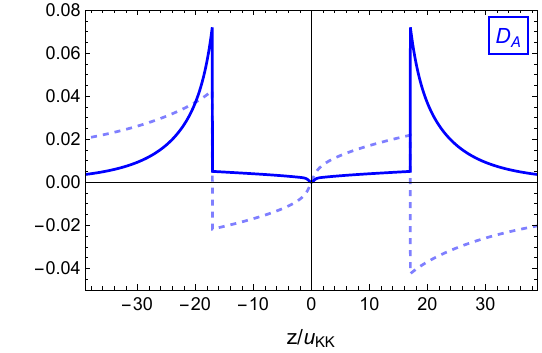}}
\caption{Charge distribution $\rho(z)$ (solid) and gauge field $h(z)$ (dashed) for all phases with up to 2 discontinuities in $h(z)$, which is suitably rescaled for illustrative purposes. 
For all plots $\lambda=7.09$, while the densities are $\bar{n}_B \simeq 0.1,0.1,1.7,0.5,3.7$ for $D^0$, $D_L$, $D_R$, $D_S$, $D_A$, respectively. A yellow background indicates that the phase has a first-order baryon onset.}  
\label{fig:profiles12}
\end{center}
\end{figure} 

\begin{figure} [t]
\begin{center}
\hbox{\includegraphics[width=0.5\textwidth]{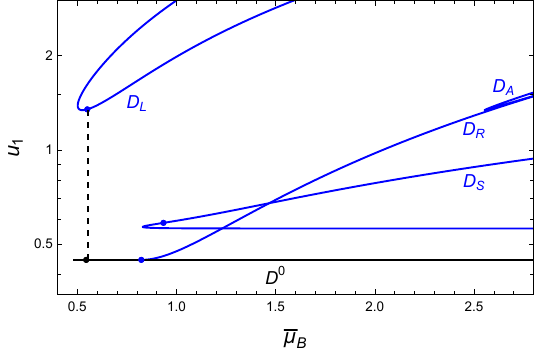}
\includegraphics[width=0.5\textwidth]{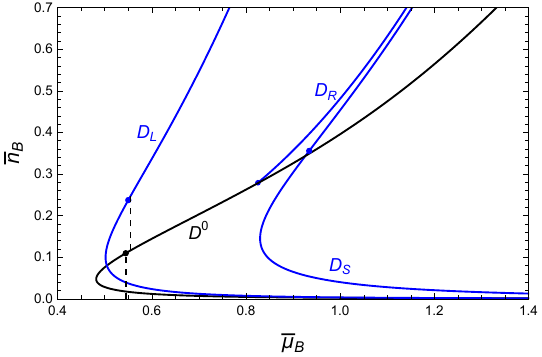}}

\includegraphics[width=0.5\textwidth]{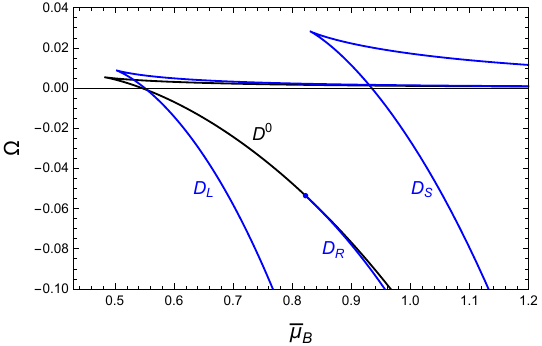}
\caption{Properties of phases with one (black) and two (blue) discontinuities, all at $\lambda = 7.09$: Location of the discontinuity $u_1$ on one half of the flavor branes (upper left panel), baryon density $\bar{n}_B$ (upper right panel), and free energy density $\Omega$ (lower panel), all as functions of the chemical potential $\bar{\mu}_B$. Dots indicate the points of the onset from the vacuum (and the point where $D_R$ connects to $D^0$). The $D_A$ configuration only exists at very large $\bar{\mu}_B$ (see upper left panel) and is outside the shown range of the two other panels. The vertical dashed lines in the upper panels indicate the first-order transitions between the vacuum and the $D^0$ phase (right panel) and between the $D^0$ and $D_L$ phases (both panels).}  
\label{fig:OMphases}
\end{center}
\end{figure}

A typical solution $h(z)$ for a fixed $\bar{n}_B$ is shown in the upper middle panel of figure \ref{fig:profiles12}, together with solutions for the $D^0$ configuration (black) and the other two-jump configurations (blue). The figure also contains the more physical charge distribution $\rho(z)$ from eq.\ (\ref{rhoz}), which is gauge invariant. Repeating the calculation for many $h_{1-}$ gives the full solutions for $u_1$, $\bar{n}_B$, and $\Omega$ as functions of $\bar{\mu}_B$, which are shown in figure \ref{fig:OMphases}. The results are shown in dimensionless units, i.e., here we do not make a choice for the value of $M_{\rm KK}$. The value for $\lambda$ used here, as in all other subsequent results that require a fixed $\lambda$, is the one from eq.\ (\ref{lamMKK}). The $D_L$ phase is similar to the $D^0$ phase in the sense that it has two branches, one of which is metastable with respect to the vacuum $\Omega=0$ and asymptotes to $\bar{\mu}_B\to\infty$ as $\bar{n}_B\to 0$. Interestingly, in both low-density and high-density limits the jump moves towards the ultraviolet, $u_1\to \infty$, as the upper left panel shows.  We also see from the lower panel that there  is a first-order transition from the vacuum to the $D_L$ phase. This transition, however, occurs in a metastable regime, i.e., at a point where the $D^0$ phase has lower free energy. Then, at $\bar{\mu}_B\simeq 0.555$, there is a first-order phase transition from $D^0$ to $D_L$.  

The existence of the $D_L$ phase  was already pointed out (not in this terminology and without comparison to the $D^0$ phase) in ref.\ \cite{Rozali:2007rx}. Since the holographic direction corresponds to an energy scale, the sharp boundary of the charge distribution was interpreted as a Fermi energy, possibly even representing a quark Fermi surface. We have not found any observable that would support this interpretation. In particular, it was later pointed out that quark degrees of freedom can be added in the form of string sources, giving rise to holographic quarkyonic matter (in combination with baryons in the pointlike approximation) \cite{Kovensky:2020xif}, and thus it seems natural to interpret all baryon number in the approach of this paper to originate from baryons, not quarks\footnote{We have checked within the deconfined geometry that string sources can also be added to the $D_R$ configuration, thus constructing a generalization of the quarkyonic phase of ref.\ \cite{Kovensky:2020xif}.}. At large $N_c$,  baryons may show fermionic as well as bosonic properties and it might be tempting to speculate whether  the different configurations we find here correspond to different fermionic and bosonic incarnations of large-$N_c$ nuclear matter. However, we have not found any consistent picture that would suggest an interpretation along these lines -- for instance interpreting certain configurations as Fermi surfaces and others as Bose-Einstein condensates -- and leave this aspect to future work. Possibly the calculation of transport properties can shed some light on the physical interpretation.

\subsection{$D_R$ phase}
\label{sec:DR}

In the $D_R$ phase, $h_{1-} = h_L(z)=0$, and the baryon density is
\be
\bar{n}_B = -\frac{3\lambda_0}{4}h_{1+}^3 \, , 
\ee
with $h_{1+}<0$. For the function $h_R(z)$ we have the initial conditions 
\be
h_R(z_1) = h_{1+} \, , \qquad \left.\frac{\partial h_R}{\partial z}\right|_{z=z_1} = \frac{2u_{\rm KK}^{1/2}\lambda_0h_{1+}^2}{3u_1^2} \, .
\ee
 Consequently, for a given $h_{1+}$, we can solve the equation of motion (\ref{EOM2}) together with the boundary condition $h_R(\infty) = 0$ for $h_R(z)$ and $z_1$. In this case, since the non-abelian gauge field is nontrivial in the ultraviolet, there is no simple expression for $\bar{\mu}_B$, which has to be computed by performing the integration in eq.\ (\ref{muB}) numerically. 

A typical profile of the $D_R$ configuration is shown in the upper right panel of figure \ref{fig:profiles12}. This figure suggests that $D_R$ can connect continuously to the $D^0$ configuration (upper left panel) by ``closing the gap'', i.e., in the limit $z_1\to 0$. In this limit,
\be \label{AcDR}
A_c = A_1(z_1\to 0) = \frac{u_{\rm KK}}{3} \, , \qquad h_{1+}\to h_c \, , 
\ee
and 
\be \label{z00}
\left.\frac{\partial h_R}{\partial z}\right|_{z_1\to 0 } = \frac{2\lambda_0h_{1+}^2}{3u_{\rm KK}^{3/2}} = \frac{h_{(1)}}{\sqrt{3u_{\rm KK}}} \, .
\ee
Using this relation for $h_{(1)}$ and eq.\ (\ref{AcDR}) for $A_c$, we see that we reproduce condition (\ref{h1Ac}). This is the stationarity condition for the jump at $z=0$ in the $D^0$ phase, i.e., the phases indeed can connect continuously. The point where $D_R$ ``grows out'' of $D^0$ can be computed by finding the point within the $D^0$ solution where the relation (\ref{z00}) is fulfilled (identifying $h_{1+}$ with $h_c$). We find that this relation is satisfied simultaneously with the $D^0$ solution at $\bar{\mu}_B\simeq 0.823$. As the lower panel of figure \ref{fig:OMphases} shows, the $D_R$ phase is energetically favored over the $D^0$ phase as soon as it appears. In contrast to the first-order transition between $D^0$ and $D_L$, this is a second-order transition, where the density changes continuously (see upper right panel of figure \ref{fig:OMphases}) and where $u_1$ increases continuously from $u_1=u_{\rm KK}$ (upper left panel). This phase transition was already observed in ref.\ \cite{CruzRojas:2023ugm}, where, at least in the WSS model, the only configuration different from $D^0$ that was considered is $D_R$. As our result in the lower panel of figure \ref{fig:OMphases} shows, this phase transition occurs in a metastable regime because the $D_L$ configuration is energetically favored  at the relevant $\bar{\mu}_B$. At least this is the case for the value of $\lambda$ chosen here; we shall discuss changes to this structure under variations of $\lambda$ in section \ref{sec:phasediagram}.

\subsection{$D_S$ phase}
\label{sec:DS}

If the jump is of the $S$ type, both functions $h_L(z)$ and $h_R(z)$ are nonzero and we need to solve the equation of motion (\ref{EOM2}) for both of them with different boundary conditions. Since the jump is symmetric we have $h_{1+} = - h_{1-}$ and we can write the baryon density as
\be
\bar{n}_B = -\frac{3\lambda_0}{2} h_{1+}^3 \, .
\ee
The initial conditions for $h_L(z)$ and $h_R(z)$ are 
\begin{subequations}
\bea
&& h_L(0) = 0 \, , \qquad \left.\frac{\partial h_L}{\partial z}\right|_{z=0} = \frac{h_{(1)}}{\sqrt{3u_{\rm KK}}} \, , \\[2ex]
&& h_R(z_1) = h_{1+} \, , \qquad \left.\frac{\partial h_R}{\partial z}\right|_{z=z_1} = \frac{2u_{\rm KK} z_1 d_{1+}}{3u_1^2}   \, .
\eea
\end{subequations}
 We need to solve the two differential equations together with the constraints 
\be
h_L(z_1) = - h_{1+} \, , \qquad \left.\frac{\partial h_L}{\partial z}\right|_{z=z_1} = \frac{2u_{\rm KK} z_1 d_{1+}}{3u_1^2} \,  , \qquad h_R(\infty) = 0 \, , 
\ee
for the functions $h_L(z)$, $h_R(z)$, and the variables $d_{1+}$, $h_{(1)}$, $z_1$, for given $h_{1+}$. We recall that the value of the abelian gauge field at $z_1$ is different for an $S$-type jump compared to all other jumps, see eq.\ (\ref{Sjump}). Therefore, the chemical potential is computed from 
\be \label{muBDS} 
\bar{\mu}_B = \frac{u_1^{5/2}\sqrt{f(u_1)}\,d_{1+}}{3\lambda_0 h_{1+}^2} + \int_{u_1}^\infty du\, \frac{\bar{n}_BQ}{u^{5/2}\sqrt{f}} \, , 
\ee
and the free energy, as usual, from the general form (\ref{Omega}). 
The lower left panel of figure \ref{fig:profiles12} shows a typical charge distribution, while $u_1$, $\bar{n}_B$, and $\Omega$ are shown in figure \ref{fig:OMphases}. Free energy and baryon density are qualitatively similar to the $D^0$ and $D_L$ phases. However, the free energy is larger than that of all other phases discussed so far and thus the $D_S$ phase is not favored for any value of the chemical potential. Interestingly, the $\bar{n}_B\to 0$ limit is realized differently from the $D_L$ phase: While in the $D_L$ phase $u_1\to \infty$, the location of the $S$-type jump approaches a finite value  as the density goes to zero. Numerically we find $u_1\to 0.56$ in this limit. In both phases, however, $\bar{\mu}_B\to \infty$, which indicates an unphysical infinite vacuum mass of the baryon. (Since the homogeneous ansatz used here cannot be expected to work down to very low densities this is not a concern for our approach.) We can identify the different behavior from the corresponding equations: in the $D_L$ phase, eq.\ (\ref{muBDL}) shows that $u_1\to \infty$ implies $\bar{\mu}_B\to\infty$. In the $D_S$ phase, we see from eq.\ (\ref{muBDS}) that $\bar{\mu}_B\to\infty$ with a finite $u_1$ is possible because, as the density goes to zero, we have $d_{1+}/h_{1+}^2\to \infty$ (both $d_{1+}$ and $h_{1+}$ go to zero).

\subsection{$D_A$ phase}
\label{sec:DA}

As for the $D_S$ phase, the $D_A$ phase requires us to solve for both nonzero functions $h_L(z)$ and $h_R(z)$. In this case, $h_{1-}$ and $h_{1+}$ are not related by a simple condition, and the density is computed from 
\be \label{nBA}
\bar{n}_B = \frac{3\lambda_0}{4}(h_{1-}^3-h_{1+}^3) \, .
\ee
Since, for an $A$-type jump, function value and derivative on either side of the jump are connected, see eq.\ (\ref{Ajump}), the initial conditions for $h_L(z)$ and $h_R(z)$ are 
\begin{subequations}
\bea
&& h_L(0) = 0 \, , \qquad \left.\frac{\partial h_L}{\partial z}\right|_{z=0} = \frac{h_{(1)}}{\sqrt{3u_{\rm KK}}} \, , \label{initialA1} \\[2ex]
&& h_R(z_1) = h_{1+} \, , \qquad \left.\frac{\partial h_R}{\partial z}\right|_{z=z_1} = \frac{2u_{\rm KK}^{1/2}\lambda_0h_{1+}^2}{3u_1^2} \, ,\label{initialA2}
\eea
\end{subequations}
and at the upper boundaries we have the constraints
\be 
h_L(z_1) =  h_{1-} \, , \qquad \left.\frac{\partial h_L}{\partial z}\right|_{z=z_1} = \frac{2u_{\rm KK}^{1/2}\lambda_0h_{1-}^2}{3u_1^2}  \,  , \qquad h_R(\infty) = 0 \, . \label{boundaryA}
\ee 
For a given $h_{1-}$, these conditions together with the equation of motion, can be solved for $h_L(z)$, $h_R(z)$ and the variables $h_{1+}$, $h_{(1)}$, $z_1$. 

Let us use this example to explain the practical calculation in some more detail, to be applied to the other cases analogously.  The system of equations can be decoupled to simplify the calculation. We can work with a given $z_1$ and first solve the differential equation for $h_L(z)$ with initial conditions (\ref{initialA1}). This requires to find the value of $h_{(1)}$ such that the two first conditions of eq.\ (\ref{boundaryA}) are met. In these conditions, $h_{1-}$ can be eliminated by inserting $h_{L}(z_1)$ directly into the condition of the derivative. Therefore, the first step is reduced to determining the single variable $h_{(1)}$, which makes it easy to ensure that all possible solutions are found. Then, $h_{1-}$ is simply extracted from the solution found for $h_L(z)$. Once we know $h_{1-}$, we can employ the equation of motion for $h_R(z)$. Recall that $h_{1-}$ enters this differential equation via the function $Q$ (\ref{Q}); conversely, $h_{1+}$ does not enter $Q$ in the equation of motion for $h_L(z)$. We determine $h_{1+}$ such that $h_R(z)$ with initial conditions (\ref{initialA2}) has the correct boundary condition given in the third relation of eq.\ (\ref{boundaryA}). Again, this is a shooting problem with a single unknown variable. The results are then used to compute $\bar{n}_B$ via eq.\ (\ref{nBA}) and we obtain $\bar{\mu}_B$ and $\Omega$ in the usual way. 

It turns out that solutions for the $D_A$ phase exist only for large $\bar{\mu}_B$ and $u_1$. We find two branches that merge at about $\bar{\mu}_B\simeq 2.55$. They are shown in the upper left panel of figure \ref{fig:OMphases} (barely distinguishable, and one branch is almost identical with the $D_R$ phase in this plot). Neither of these branches is energetically preferred and thus the $D_A$ phase plays no role for the ground state.
In the other two panels of figure \ref{fig:OMphases} the $D_A$ phase does not appear because we have chosen a smaller range in $\bar{\mu}_B$.

\subsection{Varying the coupling}
\label{sec:phasediagram}

\begin{figure} [t]
\begin{center}
\includegraphics[width=0.6\textwidth]{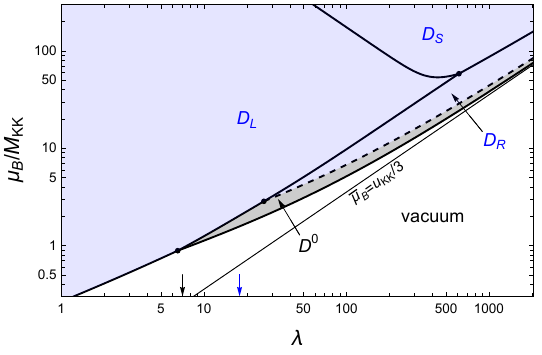}
\caption{Phase diagram in the plane of chemical potential $\mu_B$ (in units of the Kaluza-Klein mass $M_{\rm KK}$) and 't Hooft coupling $\lambda$, including phases with one and two discontinuities. Solid (dashed) curves are first- (second-) order phase transitions. The line $\bar{\mu}_B=u_{\rm KK}/3$ is added for comparison. The arrows at the bottom indicate the two values of $\lambda$ at which the $D^0$ configuration (black arrow) and the $D_L$ configuration (blue arrow) reproduce saturation density and binding energy of real-world nuclear matter. This phase diagram only serves instructive purposes since it is significantly altered by including  configurations with more than two jumps, discussed in section \ref{sec:multi}.  }  
\label{fig:2jumpphases}
\end{center}
\end{figure}

So far, all our results were obtained at the fixed coupling $\lambda=7.09$, the value for which the $D^0$ phase has realistic $n_0$ and $\mu_0$ (at a certain $M_{\rm KK}$). The exception was  figure \ref{fig:Komp}, where we have already explored the $\lambda$-dependence of the saturation properties of the $D^0$ and $D_L$ phases. To understand the dependence on the coupling better we have computed all phase transition curves between the phases discussed so far as functions of $\lambda$. The resulting phase diagram is shown in figure \ref{fig:2jumpphases}. Recall that the relation between $\bar{\mu}_B$ and $\mu_B$ contains a factor $\lambda_0$, see eq.\ (\ref{mumu}). Therefore, to avoid any confusion in a diagram where the horizontal axis is $\lambda$, we use the $\lambda$-independent dimensionless chemical potential $\mu_B/M_{\rm KK}$, not $\bar{\mu}_B$, for the vertical axis. We also keep in mind that our calculation in the classical gravity approximation is only fully accurate at infinite coupling. Therefore, exploring finite values of the coupling -- as usually done in this model and many comparable holographic applications -- is, strictly speaking, an uncontrolled extrapolation. Similar caution is required for ultra-high densities, which, in a full top-down approach, would require the use of the DBI action and the self-consistent treatment of the background geometry.

Let us first recover the result for the value of $\lambda$ used in figure \ref{fig:OMphases}, indicated by a black arrow in figure \ref{fig:2jumpphases}: there is a first-order transition from the vacuum to the $D^0$ phase, followed by a first-order transition to the $D_L$ phase. The other (blue) arrow indicates the coupling where the $D_L$ phase reproduces the correct $n_0$ and $\mu_0$, see  right panel of figure \ref{fig:Komp}. The phase diagram shows that the  baryon onset from the vacuum to the $D_L$ phase -- considered in figure \ref{fig:Komp} -- is in the metastable regime at this particular value $\lambda = 17.8$. Just like for the smaller $\lambda=7.09$, the baryon onset is to the $D^0$ phase. Hence, the reduction in stiffness at saturation through the second jump observed in figure \ref{fig:Komp} is only relevant if the $D^0$ phase is ignored. There is actually a regime, at even smaller values of the coupling, $\lambda\lesssim 6.56$, where the baryon onset {\it is} directly to the $D_L$ phase, without further phase transitions at higher densities. However, as the right panel of figure \ref{fig:Komp} shows, there is no choice of $M_{\rm KK}$ that could make the properties at saturation reasonably physical in this weak-coupling regime. 

The phase diagram shows a richer structure as we move to strong coupling. One striking feature is that for $\lambda\gtrsim 26.4$ the $D_R$ phase starts to appear. It appears through the second-order transition already seen in figure \ref{fig:OMphases}. There is also a regime at large coupling where the $D_S$ phase    has the lowest free energy. This regime is bounded by first-order transitions from the $D_L$ and from the $D_R$ phase.  The numerical evaluation becomes more challenging for extremely large values of $\lambda$ (and we have not found a simple approximation based on a large-$\lambda$ expansion of our equations). Nevertheless, the phase diagram suggests an interesting asymptotic behavior: both the $D^0$ baryon onset and the second-order onset of the $D_R$ phase seem to approach $\bar{\mu}_B=u_{\rm KK}/3$. This is the vacuum mass of the baryon in the pointlike approximation. In other words, baryonic matter in the pointlike approximation has an (unphysical) second-order onset from the vacuum at this value of $\bar{\mu}_B$ \cite{Bergman:2007wp}. Our phase diagram suggests that our finite-width solutions asymptote to pointlike solutions as the coupling goes to infinity. For the $D^0$ phase this was already pointed out in ref.\ \cite{Li:2015uea}. We will discuss the comparison and the connection to multi-layer pointlike solutions in more detail in the next section.

\section{Creating multiple instanton layers from multi-jump solutions}
\label{sec:multi}

The previous section raises the obvious question of  whether more than 2 discontinuities must be considered to find the ground state of the system for a given coupling and a given chemical potential. We have already mentioned multi-jump configurations at the end of section \ref{sec:setup} and provided the notation for such solutions.  Let us first count the number of possible configurations with up to four jumps. For 3 jumps we need to consider the configurations $D^0_X$ with $X\in \{L,R,S,A\}$, and it seems there are four possibilities. However, having a discontinuity at $z=0$ implies that $h(z)$ cannot be identically zero up to the next discontinuity at $z_1$. This is what a discontinuity of the $R$ type would require, see figure \ref{fig:LRSA}. Therefore, there are only 3 possible 3-jump configurations, namely $D^0_L$, $D^0_S$, $D^0_A$. Similarly, 4-jump solutions are denoted by $D_{XY}$ with $X,Y\in \{L,R,S,A\}$, and again some of the combinations have to be discarded: only 10 out of 16 are allowed. All possibilities, together with the 2-jump solutions from section \ref{sec:onejump}, are collected in Table \ref{tab:D}.

\begin{table}[t]
\begin{center}
\begin{tabular}{|c || c || c || c |c |c| c | |} 
 \hline
   & -- & 0 & $L$ & $R$ & $S$ & $A$ \\ [0.5ex] 
 \hline\hline
 $\;\; L\;\;$ & $\;\;D_L\;\;$ & $\;\;D_L^0\;\;$ & -- & $\;D_{RL}\;$ & $\;D_{SL}\;$ & $\;D_{AL}\;$ \\
  \hline
 $\;\; R\;\;$ & $\;\;D_R\;\;$ & -- & $\;D_{LR}\;$ & -- & -- & --   \\
 \hline
  $\;\; S\;\;$ & $D_S$ & $D_S^0$ & --  & $D_{RS}$ & $D_{SS}$ & $D_{AS}$   \\
 \hline
  $\;\; A\;\;$ & $D_A$ & $D_A^0$ & --  & $D_{RA}$ & $D_{SA}$ & $D_{AA}$   \\
 \hline
\end{tabular}
\caption{Possible configurations with 2 -- 4 discontinuities. Each column header shows the ``first'' discontinuity, either none (first column, 2-jump phases), one at $z=0$ (second column, 3-jump phases), or one at $z>0$ (third to sixth columns, 4-jump phases). The $z>0$ discontinuities are characterized by one of the four types $L$, $R$, $S$, $A$, see figure \ref{fig:LRSA}. Missing entries indicate that the respective configuration does not exist. The only configuration with a single discontinuity is $D^0$. }
\label{tab:D}
\end{center}
\end{table}

Before computing the properties of these phases, let us emphasize that classifying the phases according to the number of jumps of the non-abelian gauge field is mostly bookkeeping, not a physically meaningful classification. A more physical way of thinking about the different configurations is to consider the gauge invariant charge distribution $\rho(z)$. This is the object in which multiple instanton layers would be manifest upon considering spatially non-homogeneous solutions. To make the connection to this picture we interpret the number of local maxima of $\rho(z)$ as the number of instanton layers. As figure \ref{fig:profiles12} suggests, for 1 or 2 jumps the number of discontinuities is identical to the number of layers: $D^0$ is a 1-layer phase while $D_L$, $D_R$, $D_S$, $D_A$ are 2-layer phases. However, as we shall see, this is no longer true for configurations with 3 and 4 jumps. Moreover, even figure \ref{fig:profiles12} is somewhat deceptive because the $D^0$ phase dynamically develops additional layers at high density, i.e., it becomes a 3-layer phase without additional discontinuities. This transformation into a higher-layer phase does not occur (at all densities we checked) for the 2-jump configurations. 

Constructing solutions with 3 and 4 jumps is a relatively straightforward generalization of the 2-jump calculation and we do not explain the details of all phases. Let us instead pick two phases as examples to explain the calculation, the 3-jump phase $D_L^0$ and the 4-jump phase $D_{RL}$. As we shall see later from the numerical results, these are the two most relevant configurations.

\subsection{$D_L^0$ phase}
\label{sec:D0L}

In this phase, there is a jump at $z=0$, i.e., a nonzero $h_c<0$, and a jump at $z=z_1$ with a nonzero $h_{1-}$, while  $h_{1+}=0$. For all $z>z_1$ we have $h(z)=0$, hence we can use the notation of eq.\ (\ref{hupdown}) with $h_R(z)=0$ and have to solve the equation of motion for $h_L(z)$ with initial conditions
\be
h_L(0) = h_c \, , \qquad \left.\frac{\partial h_L}{\partial z}\right|_{z=0} = \frac{h_{(1)}}{\sqrt{3u_{\rm KK}}} \, , 
\ee
and boundary conditions for the $L$-type discontinuity exactly as in eq.\ (\ref{boundaryL}). If we fix, say, $z_1$, we need to determine the three variables $h_c$, $h_{(1)}$, $h_{1-}$. The two boundary conditions plus the differential equation (\ref{EOM2}) with the two initial conditions do not give enough constraints. The additional equation needed is the one obtained from the continuity of $A(z)$. In the case of the $D^0_L$ phase, this is eq.\ (\ref{A0Ac}). Inserting $A_1$ from eq.\ (\ref{Ljump}) and $A_c$ from eq.\ (\ref{h1Ac}), this reads
\be
\frac{u_1}{3} = \frac{u_{\rm KK}^2h_{(1)}}{2\sqrt{3}\lambda_0h_c^2}-
\frac{\lambda_0 h_c^3}{2u_{\rm KK}^{3/2}}\arctan\frac{z_1}{u_{\rm KK}} 
+ \frac{3\lambda_0}{4}\int_{u_{\rm KK}}^{u_1} du\frac{h_L^3}
{u^{5/2}\sqrt{f}} \, .
\ee
Now the system of equations can be solved, and we can compute the chemical potential from the same equation as in the $D_L$ phase, see eq.\ (\ref{muBDL}). 

\begin{figure} [t]
\hbox{\includegraphics[width=0.33\textwidth]{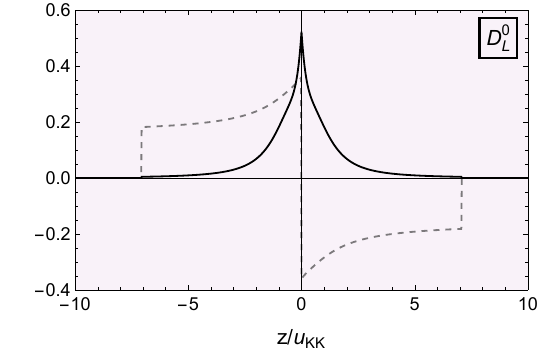}
\includegraphics[width=0.33\textwidth]{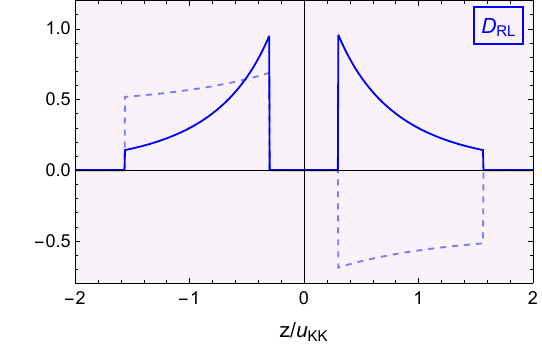}\includegraphics[width=0.33\textwidth]{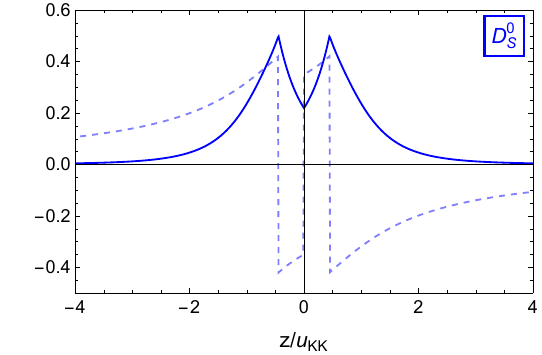}}
\hbox{\includegraphics[width=0.33\textwidth]{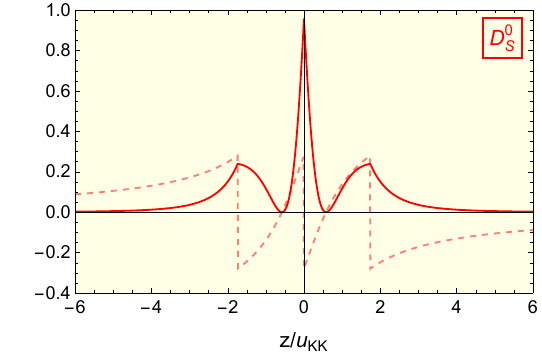}\includegraphics[width=0.33\textwidth]{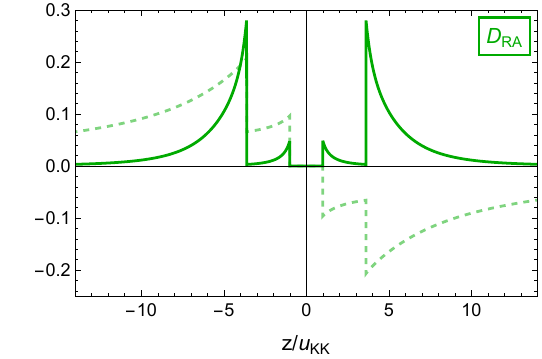}\includegraphics[width=0.33\textwidth]{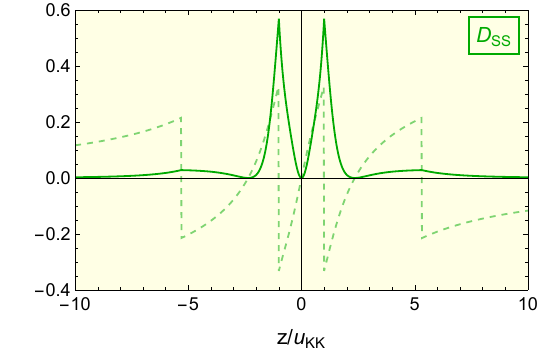}}
\caption{Selection of profiles with more than two jumps, resulting in 1 (black), 2 (blue), 3 (red), or 4 (green) local maxima (``instanton layers''). As in figure \ref{fig:profiles12}, $\lambda=7.09$, and the vertical scales are  valid for the normalized charge distribution $\rho(z)$ (solid curves), while the functions $h(z)$ (dashed curves) have been multiplied with convenient factors. Configurations with shaded background have a zero-density limit, with purple (yellow) background indicating 
finite (infinite) vacuum mass of the baryons. The densities are, from left to right, $\bar{n}_B \simeq 0.3,1.1,0.3$ (upper row) and $\bar{n}_B \simeq 0.03,4.0,1.5$ (lower row). }
\label{fig:profileLayers2}
\end{figure}

In contrast to the phases with one or two jumps, the $D_L^0$ configuration has distinct classes of solutions. The reason is that $h_L(z)$ may or may not have a node, i.e., with $h_c<0$ it turns out that there are solutions for both $h_{1-}<0$ and $h_{1-}>0$. The profile of the former, which is the physically more interesting solution for reasons to be explained momentarily, is shown in the upper left panel of figure \ref{fig:profileLayers2}. (The same structure of solutions exists for many other configurations; the figure does show both solutions -- with and without node -- for the $D_S^0$ configuration.) We see that although we have three discontinuities in the gauge field, the charge distribution only has a single maximum, which we interpret as a single instanton layer. Just like the $D^0$ phase, and as can be anticipated from the barely visible shoulder-like structure in $\rho(z)$, the $D^0_L$ phase also turns dynamically into a 3-layer phase at large densities.

\subsection{$D_{RL}$ phase}
\label{sec:DRL}

In the $D_{RL}$ phase, $h(z)$ is only nonzero in the domain $z\in[z_1,z_2]$. The boundary conditions at both boundaries have the same structure,  
\be
h(z_1) = h_{1+} \, , \qquad \left.\frac{\partial h}{\partial z}\right|_{z=z_1} = \frac{2u_{\rm KK}^{1/2}\lambda_0h_{1+}^2}{3u_1^2} \, ,
\ee
\be
h(z_2) = h_{2-} \, , \qquad \left.\frac{\partial h}{\partial z}\right|_{z=z_2} = \frac{2u_{\rm KK}^{1/2}\lambda_0h_{2-}^2}{3u_2^2} \, .
\ee
Additionally, we have condition (\ref{Acont}) with $k=2$. With the $L$- and $R$-type relation $A_k=u_k/3$, this condition becomes
\be
\frac{u_2}{3} = \frac{u_1}{3} + \frac{3\lambda_0}{4}\int_{u_1}^{u_2} du\frac{h^3}{u^{5/2}\sqrt{f}}-\frac{\lambda_0 h_{1+}^3}{3u_{\rm KK}^{3/2}} \left(\arctan\frac{z_2}{u_{\rm KK}} -\arctan\frac{z_1}{u_{\rm KK}}\right) \, .
\ee
Now, for given $z_1$, we can solve the system of equations for the function $h(z)$ and the three variables $h_{1+}$, $h_{2-}$, $z_2$.  
As in all cases we consider, and analogous to the details explained in the context of the $D_A$ phase, see Sec.\ \ref{sec:DA}, the system of equations can be decoupled. In this case it is two variables at most we have to solve for simultaneously, simple enough to systematically check for all possible solutions. Since this phase is trivial in the ultraviolet (like all phases whose sequence of jumps ends with an $L$-type jump), the chemical potential has the simple form (\ref{muBDL}) with $u_1$ and $z_1$ replaced by $u_2$ and $z_2$.

The profile of the $D_{RL}$ phase is shown in the upper middle panel of figure \ref{fig:profileLayers2}. We interpret this solution as a two-layer phase due to the two local maxima of $\rho(z)$. Just like for the $D^0_L$ configuration, there is a second, less relevant, solution, where $h(z)$ has a node and where we find 4 layers.

\subsection{Connecting to pointlike configurations}
\label{sec:pointlike}

\begin{figure} [t]
\begin{center}
\includegraphics[width=0.5\textwidth]{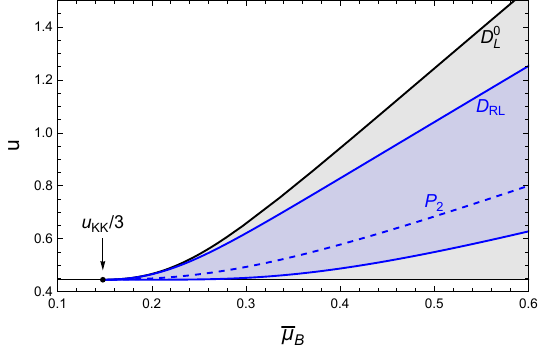}

\hbox{\includegraphics[width=0.33\textwidth]{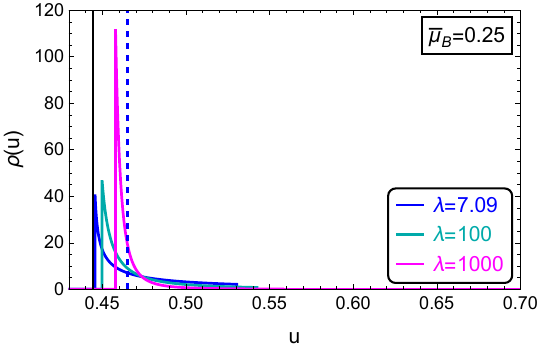}\includegraphics[width=0.33\textwidth]{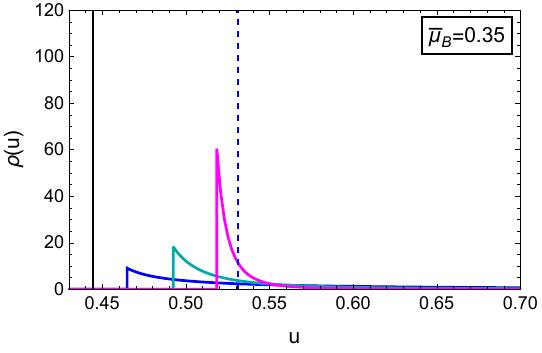}\includegraphics[width=0.33\textwidth]{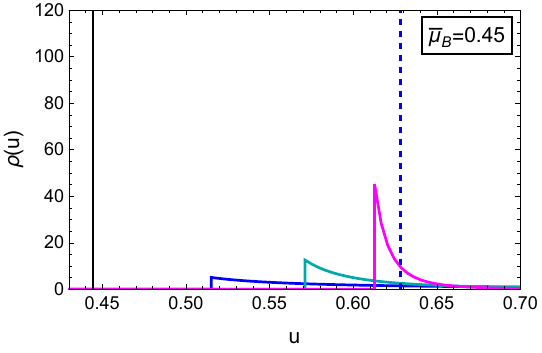}}
\caption{{\it Upper panel:} Region of nonzero charge distribution $\rho(u)$ on one half of the flavor branes, for the $D^0_L$ phase (black, between $u_{\rm KK}$ and $u_1$) and the $D_{RL}$ phase (blue, between $u_1$ and $u_2$), all as a function of the chemical potential and for $\lambda=7.09$. The blue dashed line is the location of the delta-peak in the 2-layer pointlike phase $P_2$ (in the $P_1$ phase the baryons always sit at $u_{\rm KK}$). The plot demonstrates that both $D_L^0$ and $D_{RL}$ become pointlike for $\bar{\mu}_B\to u_{\rm KK}/3$. {\it Lower panels:} Charge distributions $\rho(u)$ of the $D_{RL}$ phase for three different values of $\bar{\mu}_B$ and three different coupling strengths. The vertical dashed line indicates the location of the baryons in the $P_2$ phase, and the thin vertical solid line marks $u=u_{\rm KK}$. These plots indicate that the $D_{RL}$ configuration becomes pointlike for $\lambda\to \infty$.}  
\label{fig:width}
\end{center}
\end{figure}

The two phases just discussed, $D_L^0$ and $D_{RL}$, are of physical interest for several reasons. Firstly, as we go to lower densities, we find that the width of their charge distribution $\rho(z)$ becomes smaller and smaller. In the $D_L^0$ phase this means that $z_1\to 0$ as $\bar{n}_B\to 0$. In the $D_{RL}$ phase we find that both $z_1,z_2\to 0$, i.e., while the width becomes smaller, the baryons also move towards the tip of the connected flavor branes. Moreover, in both phases, as the density goes to zero the chemical potential approaches a finite and nonzero value, $\bar{\mu}_B\to u_{\rm KK}/3$, i.e., these configurations reproduce a finite vacuum mass of the baryon. This is different from all phases discussed so far, which either approached $\bar{\mu}_B\to\infty$ ($D^0$, $D_L$, $D_S$) or did not have a zero-density limit ($D_R$, $D_A$), see figure \ref{fig:OMphases}. In fact, $D_L^0$ and $D_{RL}$ are the only phases among all  configurations of Table \ref{tab:D} that have these  properties. These properties -- becoming pointlike at low densities and reproducing the vacuum mass of a single baryon -- are manifest in the upper panel of figure \ref{fig:width}. This figure shows the domain $u\in[u_{\rm KK},u_1]$ for the $D^0_L$ phase (black) and the domain $u\in[u_1,u_2]$ for the $D_{RL}$ phase, as functions of the chemical potential.  

The second-order onset at $\bar{\mu}_B=u_{\rm KK}/3$ is also obtained if the charge distribution is assumed to be pointlike to begin with, as already mentioned in the context of figure \ref{fig:2jumpphases}. We present the calculation of pointlike configurations for an arbitrary number of layers in appendix \ref{app:pointlike} and refer to them as 
\be
P_\ell: \quad \mbox{configuration with $\ell$ pointlike baryon sources.} 
\ee
The location of the baryons in the $P_2$ configuration\footnote{First considered in ref.\ \cite{Kovensky:2020xif} in the deconfined version of the model.} is shown in figure \ref{fig:width} (blue dashed curve). The location of the baryons in the $P_1$ phase is, by symmetry, at $u=u_{\rm KK}$ for all chemical potentials. 

Our calculation thus describes how to smooth out the simplest two of these configurations by providing the 1-layer ($D_L^0$) and 2-layer ($D_{RL}$) generalizations 
of the corresponding pointlike approximations within a spatially homogeneous ansatz for the gauge fields. In the previous literature these solutions have always been treated as distinct approximations and connections between them were at best observed indirectly in the $\lambda\to\infty$ limit. (The pointlike limit is recovered more directly from a finite-width instanton ansatz \cite{Li:2015uea}.) Here our configurations connect to the pointlike solutions by going to low densities (at fixed coupling). Additionally, we may ask whether our finite-width configurations also become pointlike as the coupling is increased at fixed nonzero density. This is expected from the discussion of the phase diagram in figure \ref{fig:2jumpphases}. It can also be  expected from single baryons sitting at $z=0$, whose width goes to zero as $\lambda\to\infty$. Indeed, we find that this is the case, as the lower panels of figure \ref{fig:width} show for the $D_{RL}$ phase (an analogous result can be obtained for the $D_L^0$ phase). They show the charge distribution\footnote{In slight abuse of notation, we denote the charge density per unit length in $u$ by $\rho(u)$, which differs by a factor $dz/du$ from the charge density $\rho(z)$ per unit length in $z$, ensuring the normalization $\int_{u_{\rm KK}}^\infty du\, \rho(u) = \frac{1}{2}$.}
$\rho(u)$ for three different chemical potentials and three different values of $\lambda$. Going from high density (right panel) to low density (left panel) shows that the width of the distribution decreases, as already seen in the upper panel. Now, additionally, we see in each panel separately that the distribution becomes narrower as the coupling is increased while, also, the location of the maximum approaches the location of the pointlike baryon layer. All distributions are nonzero in the finite domain $u\in[u_1,u_2]$, but this figure shows that, for large couplings, the baryon charge is concentrated in a region much smaller than this domain.

\subsection{Free energy comparison and incompressibility}
\label{sec:compare}

\begin{figure} [t]
\begin{center}
\hbox{\includegraphics[width=0.5\textwidth]{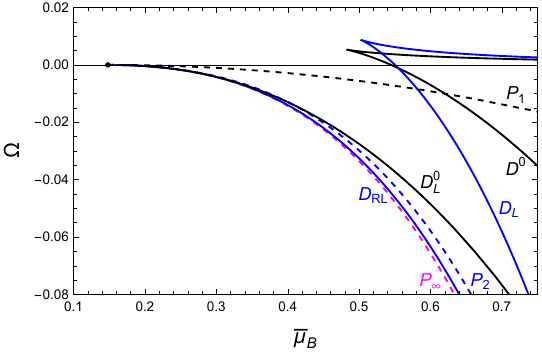}\includegraphics[width=0.5\textwidth]{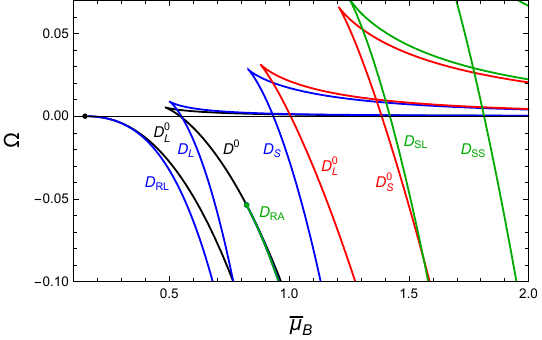}}
\caption{{\it Left panel:} Free energy density as a function of the chemical potential of the most relevant phases with up to 4 discontinuities, resulting in 1 (black) and 2 (blue) layers. Dashed curves are the free energies of the pointlike phases $P_1$ and $P_2$ and the continuum limit of infinitely many pointlike layers $P_\infty$ (magenta). {\it Right panel:} Free energy over a larger range, including all phases with 3 (red) and 4 (green) layers that have a first-order onset from the vacuum. The curve labeled by $D_{RA}$ is indistinguishable on the scale shown here from two additional curves, i.e., there are three phases that connect continuously to the $D^0$ curve at the same point (within our numerical uncertainty); namely, in order of decreasing free energy, $D^0_S$ (2-layer version, see top right panel of figure \ref{fig:profileLayers2}), $D_R$ (see figure \ref{fig:OMphases}), and $D_{RA}$.   }  
\label{fig:omega2}
\end{center}
\end{figure}

The two phases $D_{RL}$ and $D^0_L$ are also of particular interest because they are the energetically favored solutions (unless the density is very high). This is shown with the help of the free energy density in figure \ref{fig:omega2}. For all values of $\bar{\mu}_B$ shown in the left panel, the $D_{RL}$ phase has the lowest free energy of all configurations in Table \ref{tab:D}. This panel also shows the comparison to the pointlike approximation. For 1 and 2 layers, allowing  each baryon layer to spread out in the holographic direction reduces the free energy: the 1-layer $D_L^0$ (2-layer $D_{RL}$) phase has lower free energy for all chemical potentials than its 1-layer $P_1$ (2-layer $P_2$) pointlike counterpart. Using the results of appendix \ref{app:pointlike}, we find that the free energy of the pointlike phases decreases monotonically with the number of layers for any given $\bar{\mu}_B$ and that it asymptotes to a finite result as the number of layers goes to infinity. Appendix \ref{app:pointlike} also contains the derivation of semi-analytical expressions for this continuum limit $P_\infty$. Interestingly, the charge distribution of the $P_\ell$ solution asymptotes to a continuous 2-layer form according to our classification in terms of local maxima, see figure \ref{fig:rhoInf} in appendix \ref{app:pointlike}. We observe that this $P_\infty$ configuration is (just about) energetically preferred over the $D_{RL}$ solution. 

Does this observation depend on the value of the coupling $\lambda$? The left panel of figure \ref{fig:omega2} compares the $D_{RL}$ result for $\lambda=7.09$ with the pointlike results, which are invariant under changes of $\lambda$ in the dimensionless units used here. We have checked that, going down in $\lambda$, the free energy of the $D_{RL}$ phase becomes smaller but does not seem to decrease below the magenta dashed curve. (As we {\it increase} $\lambda$, the $D_{RL}$ curve approaches the 2-layer pointlike result, i.e., goes up in free energy, as expected from the discussion in the previous subsection.) The numerical results thus suggest that $P_\infty$ is the preferred phase at low densities for all $\lambda$, at least compared to the finite-width solutions discussed so far. 

Is there any multi-jump, finite-width configuration that is preferred over $P_\infty$? Let us first discuss the phases with up to 4 jumps listed in Table \ref{tab:D}. The right panel of figure \ref{fig:omega2} shows all configurations we have found for the given $\Omega$-$\bar{\mu}_B$ domain of that plot. In particular, it shows all phases with a first-order baryon onset (i.e., the phases whose free energy curves intersect with $\Omega=0$). We observe  the tendency that phases with higher number of layers become less relevant, in contrast to the pointlike approximation where we have observed the opposite trend. Two of the phases with a first-order baryon onset and more than two layers are illustrated in figure \ref{fig:profileLayers2} by their charge distributions, namely the  $D_S^0$ phase (3 layers) and the $D_{SS}$ phase (4 layers). The right panel of figure \ref{fig:omega2} shows another interesting feature: We recall from figure \ref{fig:OMphases} that the $D_R$ phase connects continuously to the simple 1-layer $D^0$ phase. We now find that two more phases connect at the same point (within our numerical uncertainty) to the same branch, namely a second, 2-layer, configuration of $D_S^0$ (profile shown in the upper right panel of figure \ref{fig:profileLayers2}) and the 4-layer $D_{RA}$ configuration (also shown in figure \ref{fig:profileLayers2}, see lower middle panel). On the scale of figure \ref{fig:omega2}, the curves of all three phases are indistinguishable, but we have checked that the $D_{RA}$ phase has the lowest free energy among them. 

Several of the phases in Table \ref{tab:D} neither have a first-order baryon onset nor are they competitors for the ground state at any density, at least for the value of the coupling $\lambda=7.09$ used for figure \ref{fig:omega2}. One  phase not discussed so far, however, plays a role at this coupling strength at larger densities, beyond the range shown in figure \ref{fig:omega2}. This is the (4-layer) $D_{AL}$ phase, which only exists at nonzero densities, similar to the $D_A$ phase in figure \ref{fig:OMphases}. Ignoring any pointlike configurations for a moment, we find that there is a first-order transition from $D_{RL}$ to $D_{AL}$ at $\bar{\mu}_B\simeq 1.37$, and a subsequent second-order transition from $D_{AL}$ to $D_L$ at $\bar{\mu}_B\simeq 2.94$. These values are only valid for the particular coupling strength used here. We have not attempted to extend this observation to different couplings and produce a phase diagram analogous to the 2-jump phase diagram in figure \ref{fig:2jumpphases}. The results suggest, however, that the $D_{RL}$ configuration will be the preferred phase among all our finite-width solutions at not too large densities for all $\lambda$, because the onset of this phase always occurs at the point where $\bar{\mu}_B$ equals the value of the baryon mass $u_{\rm KK}/3$, which is the lowest possible onset point.  

Coming back to the comparison with the pointlike phases, we have thus seen that none of our solutions with up to 4 jumps is favored over the $P_\infty$ configuration. How about more than 4 jumps? Our results give us some guidance on how to search for such a phase. Most naturally, we may ask whether there is any phase of the type $D_{RLRL\ldots RL}$, consisting of ``$RL$ blocks'' with $h(u)=0$ between them. Interestingly, such phases with more than one $RL$ block on each half of the flavor branes do not exist, as one can show as follows. Suppose we are looking for a $D_{RLRL}$ configuration, with discontinuities at $z_1<z_2<z_3<z_4$. The key observation is that the equations for the ``first'' $RL$ block between $z_1$ and $z_2$ decouple from the ones for the second block. More precisely, for a given $z_1$ we can solve for $h(u)$ of the first block, including the values for $h_{1+}$, $h_{2-}$, $z_2$, see section \ref{sec:DRL}. Then, the continuity condition for $A(z)$ (\ref{Acont}) between $z_2$ and the lower end of the second block, $z_3$, yet to be determined, reads   
\be
\frac{u_3}{3}-\frac{u_2}{3} =  \int_{u_2}^{u_3} du\, A' \, , 
\ee
which can be written in the form
\be
\alpha(z_2) = \alpha(z_3) \, , \qquad \alpha(z) \equiv \frac{u(z)}{3}+\frac{3\lambda_0}{2
u_{\rm KK}^{3/2}}(h_{1+}^3-h_{2-}^3)\arctan \frac{z}{u_{\rm KK}} \, .
\ee
 For this simplification, $h(z)=0$ for $z\in[z_2,z_3]$ is crucial. With $h_{2-}>h_{1+}$ (see figure \ref{fig:profileLayers2}) the function $\alpha$ is indeed not one-to-one and thus in principle there might exist a $z_3$ to fulfil $\alpha(z_2) = \alpha(z_3)$. In order to find this $z_3$ we may go along the branch of $D_{RL}$ solutions and numerically check for the condition $\alpha(z_2) = \alpha(z_3)$. We find numerically that along the entire branch there is no point that allows for a $z_3>z_2$ and thus we conclude that the equations of motion together with our stationarity conditions do not allow to add a second $RL$ block to the $D_{RL}$ configuration. 

This argument excludes more than just the $D_{RLRL}$ phase: It excludes all phases with arbitrarily many $RL$ blocks, i.e., all phases $D_{RLRL\ldots RL}$. Even more generally, it says that if we start with an $RL$ block in $z\in[z_1,z_2]$ we need to keep $h(z)=0$ for all $z>z_2$. Any additional sequence of jumps would have to start with a jump of the $R$ type, and this is not allowed as we have just argued. Consequently, the most straightforward search for a multi-layer phase with layers of finite widths does not yield any new phases. 
This observation is reminiscent of previous results, which have concluded within an instantonic ansatz \cite{Preis:2016fsp} and a homogeneous ansatz on the level of the field strengths \cite{Elliot-Ripley:2016uwb}, that phases with more than 2 layers are not favored. We are left with the interesting result that, as far as the present study goes, the most preferred configuration is the one with infinitely many pointlike layers; of course, without having checked all possible multi-jump configurations, so this is only a rigorous result for up to 4 jumps plus the observation of the absence of $D_{RLRX}$ phases, where $X$ is any sequence of $L$, $R$, $S$, $A$ jumps.

\begin{figure} [t]
\begin{center}
\hbox{\includegraphics[width=0.5\textwidth]{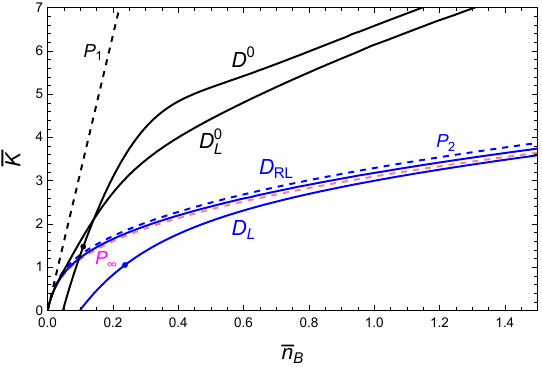}\includegraphics[width=0.5\textwidth]{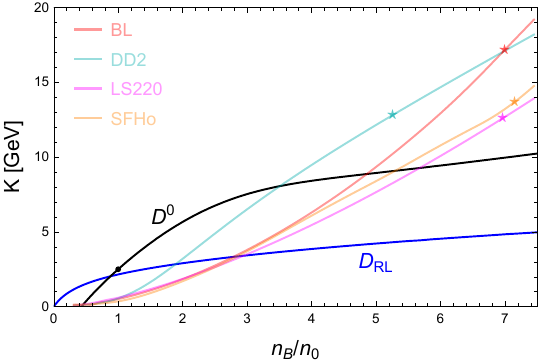}} 
\caption{{\it Left panel:} Dimensionless incompressibility for the energetically most preferred 1-layer (black, $D^0$ and $D_L^0$) and 2-layer (blue, $D_L$ and $D_{RL}$) phases, as a function of baryon density, for $\lambda=7.09$. For the phases which have a first-order baryon onset, $D^0$ and $D_L$, the onset point is marked by a dot. The dashed lines are the results for the pointlike 1-layer (black), 2-layer (blue), and $\infty$-layer (magenta) phases. The energetically favored phase is $P_\infty$, closely followed by $D_{RL}$, see figure \ref{fig:omega2}. {\it Right panel:} Incompressibility of the $D^0$ and $D_{RL}$ phases, computed with $M_{\rm KK}=1000\, {\rm MeV}$, as a function of baryon density in units of the physical saturation density $n_0$. The background shows, for comparison, the incompressibility (of beta equilibrated, charge neutral matter) computed from various non-holographic models of nuclear matter (BL, DD2, LS220, SFHo), with asterisks indicating the central point of the heaviest stable star for each model; taken from ref.\ \cite{Perego:2021mkd}, where also references for the individual models can be found.}  
\label{fig:KnB}
\end{center}
\end{figure}

Finally, let us return to the stiffness of our holographic baryonic matter. In the left panel of figure \ref{fig:KnB} we plot the dimensionless incompressibility $\bar{K}$ as a function of the density $\bar{n}_B$ for the energetically most preferred one-layer and two-layer phases, including their pointlike counterparts, plus the result for the $P_\infty$ phase. The dimensionless incompressibility is defined by replacing $\mu_B$ and $n_B$ in the definition (\ref{Kdef}) by $\bar{\mu}_B$ and $\bar{n}_B$, i.e., the physical value $K$ can be obtained from $\bar{K}$ in the same way as given for the chemical potential in eq.\ (\ref{mumu}). The observations from this  panel  are as follows.

\begin{itemize}
\item The phases with a second-order baryon onset (including all pointlike phases) have $\bar{K}\to 0$ as the density goes to zero. The phases with a first-order onset ($D^0$ and $D_L$) show $\bar{K}\to 0$ at a finite density because they contain an unstable branch where the incompressibility is negative.

\item As already suggested by figure \ref{fig:Komp}, adding more jumps in $h(u)$ and indeed more instanton layers softens nuclear matter: All 2-layer phases (blue) are more compressible than all 1-layer phases (black) (except for a small low-density regime where the metastable $D^0$ branch is softer than the $D_{RL}$ and pointlike phases). 

\item The finite-width 1-layer phases -- mostly $D^0$ but in a less pronounced way also $D_L^0$ -- show a shoulder-like feature in the compressibility. This coincides with the dynamical appearance of two additional layers in the vicinity of this feature, i.e., turning from 1 to 3 layers within the same solution also appears to soften nuclear matter.

\item The 2-layer pointlike phase and indeed all $n$-layer pointlike phases for $n\ge 2$ yield a good approximation for the incompressibility of the finite-width $D_{RL}$ phase. 

\end{itemize}

To put our results into a more physical context, we show the dimensionful incompressibility in the right panel of figure \ref{fig:KnB}. Here, the $D^0$ phase, used for constructing neutron stars in ref. \cite{Kovensky:2021kzl}, together with the energetically  preferred $D_{RL}$ phase are shown. To translate the curves of the left panel into physical units we have used $M_{\rm KK}=1000\, {\rm MeV}$ for both curves, which gives the correct saturation density $n_0$ for the $D^0$ phase. Our results are compared to non-holographic models for cold nuclear matter taken from ref.\ \cite{Perego:2021mkd}. These curves are valid for matter under neutron star conditions, i.e., in equilibrium with respect to the electroweak interactions and imposing electric charge neutrality, while our curves describe isospin-symmetric nuclear matter. Since we are interested only in a qualitative comparison, this is sufficient for our purposes. The observations from the right panel are: 
\begin{itemize}

\item As discussed in section \ref{sec:D0}, we confirm that the incompressibility at saturation $n_B=n_0$ of the $D^0$ phase is too large compared to empirical data, to which the non-holographic curves are fitted. (Since the $D_{RL}$ configuration does not have a saturation density its incompressibility at $n_B=n_0$ is not particularly meaningful.) 

\item Even our stiff (and energetically disfavored) $D^0$ phase becomes softer at high densities than all non-holographic equations of state for pure nuclear matter. Perhaps this explains why realistic neutron stars are possible \cite{Kovensky:2021kzl} despite the unrealistic stiffness at saturation.

\item There is a clear difference in tendencies between holographic and non-holographic incompressibilities. The non-holographic models show a strong increase in stiffness as a function of density, while the holographic results suggest an almost constant stiffness at high densities. It is the high-density region where our approximation is most trustworthy  -- and the non-holographic nuclear matter results become unreliable.

\end{itemize}

\section{Discussion}
\label{sec:discussion}

Let us discuss significance and applicability of our findings. We turned  a physical first-order nuclear matter onset of the previous literature ($D^0$ phase) into an apparently unphysical second-order onset by allowing for more general solutions within the same homogeneous ansatz. This seems discouraging at first sight for future applications. Of course, since the WSS model is not dual to real-world QCD -- although it shares many features with it -- there is no guarantee that properties of nuclear matter are correctly reproduced. We should also recall that our homogeneous ansatz cannot be expected to be a good approximation at low densities, and it is the regime of low densities that is probed in the vicinity of the second-order onset. Therefore, although discouraging from a phenomenological point of view, this aspect of our results should not be treated as a rigorous prediction of the model. In fact, in a more fundamental instanton approach and depending on how instanton interactions are included, the model can provide a physical first-order onset of isospin-symmetric nuclear matter \cite{Preis:2016fsp}, in a phenomenological treatment of the instanton deformation even with the correct incompressibility at saturation \cite{BitaghsirFadafan:2018uzs}. It should also be mentioned that pure neutron matter does show a second-order baryon onset and is a better approximation to matter in the interior of neutron stars than isospin-symmetric matter. Therefore, for future applications our novel $D_{RL}$ configuration, once extended to isospin-asymmetric matter, is still a potential candidate phase to describe neutron and neutron star matter. Furthermore, it appears to be particularly suited to compute properties beyond thermodynamics because it might facilitate the calculation of transport properties due to its trivial behavior in the infrared and ultraviolet.  

Similarly, we may ask whether the softening of nuclear matter by allowing the location of the discontinuities to adjust dynamically is a step towards more realistic holographic nuclear matter. If we simply take our energetically favored multi-jump solution $D_{RL}$, we can no longer make the comparison to the empirical value of the incompressibility at saturation, because there is no saturation density in the $D_{RL}$ phase. Given that our approximation is best  at large (but not too large) densities, the more sensible question is if our results can be used as a prediction for the stiffness of strongly coupled nuclear matter in that high-density regime, reached in the core of neutron stars. While the $D^0$ phase (extended to isospin-asymmetric matter and after adding a  crust) does yield realistic neutron stars, this is not the case for the $P_1$ and $P_2$ phases, at least using the DBI action (as opposed to our YM approximation) \cite{Kovensky:2021kzl}. Since the incompressibility of the $D_{RL}$ phase is very close to that of the $P_2$ phase at all densities it is questionable whether realistic stars can be built with our new solution. However, the dependence of the $D_{RL}$ phase on the coupling $\lambda$ is different from the trivial scaling of the $P_2$ phase. Therefore, it needs to be checked in an actual calculation whether a suitable $\lambda$ can be found to reproduce realistic neutron stars with the $D_{RL}$ configuration (or possibly using the $P_\infty$ phase). And, of course, employing our novel phases for, e.g., qualitative strong-coupling predictions of transport properties of nuclear matter might anyway be insightful, taking into account the limitations of the model and the view that thermodynamic properties meeting all empirical constraints are not a mandatory pre-condition.  

Finally, let us also discuss the significance of the pointlike solutions. In particular, one might argue that the continuum ``2-layer'' limit $P_\infty$ is, in some sense, the main result of the paper because this phase is energetically favored over all our multi-jump solutions. This observation is remarkable but has to be treated with care for several reasons. Firstly, we cannot claim to have done an exhaustive or even systematic study of multi-jump phases with more than 4 discontinuities of the non-abelian gauge field. More importantly, the pointlike approximation contains a singular ansatz without a corresponding solution of the equations of motion, with a pointlike source whose only direct coupling is to the abelian gauge field. Therefore, the $P_\infty$ phase might provide important clues on how to construct an actual solution on a more fundamental level, but its comparison to our multi-jump phases that do solve the equations of motion is, albeit instructive, not conclusive. Despite the different ansatz, it is striking that the charge distribution of the $P_\infty$ phase looks very similar to that of some of our multi-jump configurations and one may ask whether we have missed a certain solution that reproduces the $P_\infty$ phase. However, as just argued, there is no guarantee that the pointlike ansatz reproduces any actual solutions to the equations of motion. It is thus not surprising that even its continuum limit, where the charge distribution becomes piecewise smooth, yields a configuration that cannot be reproduced within our ansatz on the level of the gauge fields. Nevertheless, the novel $P_\infty$ phase is interesting for future applications as well, not only because of its low free energy but also due to its mathematical simplicity, which reduces the numerical effort to solving a simple algebraic equation. On the other hand, since it is not built  upon an actual solution of the equations of motion it seems not straightforward to compute, e.g., transport properties in this phase.

\section{Summary and outlook}
\label{sec:summary}

We have generalized and improved the description of holographic  baryonic matter in the Witten-Sakai-Sugimoto model. This was done within the homogeneous ansatz for the non-abelian gauge fields used in previous studies, which requires a discontinuity to create topological baryon number via the CS term. We have allowed this discontinuity to be dynamical, i.e., to move as a function of density, and further considered  configurations with more than one  discontinuity. 

We have chosen to work in the simplest version of the model, employing the confined geometry, antipodal separation of the flavor branes, and the YM approximation of the gauge field action. By finding the stationary points of the effective potential with respect to the parameters that characterize the discontinuities, we have identified four different shapes that are possible at each jump. Combining them gives rise to a plethora of phases, several of which are thermodynamically relevant and cannot be ignored. We have studied these combinations systematically up to 4 discontinuities, and also provided an argument why the most straightforward extension to configurations with more discontinuities does not simultaneously satisfy the equations of motion and the stationarity conditions.

We have pointed out that multi-jump solutions can be interpreted as multi-layer instanton solutions,  the number of jumps in the gauge field not necessarily being identical to the more physical number of instanton layers. Our main results concern the 1-layer phases termed $D^0$ (from previous literature), $D^0_L$ (a 3-jump configuration), and the 2-layer phases $D_L$ (2 jumps) and $D_{RL}$ (4 jumps). Computing the incompressibility of these phases, we have found that matter from 2-layer structures is significantly softer than from 1-layer structures. Moreover, we have identified the $D_{RL}$ phase as the one with the lowest free energy among all our  multi-jump configurations. The charge distribution of the $D_{RL}$ phase has a block-like structure in the bulk with vanishing non-abelian gauge field in the infrared and the ultraviolet, which renders the numerical treatment of this phase relatively simple, and thus provides a valuable starting 
point for more complicated calculations in the future. 

We have also established a continuous connection between the $D_{RL}$ and $D^0_L$ phases and the 1- and 2-layer pointlike configurations, both in the low-density and  strong-coupling limits. At fixed coupling, we have found that our finite-width solutions lower the free energy of the system compared to forcing the charge distribution to be pointlike. Interestingly, we have also found that as the number of pointlike layers increases, a (piecewise) smooth and finite limit emerges, which we have termed the $P_\infty$ phase. Its free energy is lower than that of any finite number of delta peaks. The $P_\infty$ phase is even preferred (albeit only by a very small difference in free energy) over the $D_{RL}$ phase, the energetically favored phase among all finite-width, multi-jump solutions to the equations of motion we have found. 

Our results can be extended along several directions in future work. Perhaps most obvious, having in mind applications to neutron stars,  is the extension to isospin-asymmetric matter, which was already done with the simpler $D^0$ configuration. It would  be interesting to see if neutron and proton contributions to the baryon density can be assigned to different ``blocks'' in the bulk within the $D_{RL}$ configuration. And, indirectly, this extension might also give rise to isospin-asymmetric matter in the pointlike approximation. It would also be interesting to study pure neutron matter, given that we have found a second-order baryon onset in the $D_{RL}$ phase, which is unphysical for isospin-symmetric QCD matter. Another promising direction is the study of quarkyonic matter, which was considered in the WSS model within the pointlike approximation for baryons \cite{Kovensky:2020xif}. The discovery of the $D_{RL}$ configuration is crucial for improving the description of holographic quarkyonic matter because it allows for a straightforward generalization of ref.\ \cite{Kovensky:2020xif}, by adding string sources at the tip of the flavor branes within the deconfined geometry. The novel $P_\infty$ phase also deserves further studies, for instance by repeating our calculation in the deconfined geometry, possibly with non-antipodal flavor branes, in order to investigate its temperature dependence. 

It would also be of interest to check if configurations analogous to the ones developed in this paper also arise in other holographic models. Indeed, it is usually the case that bottom-up constructions such as V-QCD and hard-wall models share many features with the WSS model. For instance, the study of multi-jump solutions in V-QCD was initiated recently in \cite{CruzRojas:2023ugm}. It is important to check if phases which might have been overlooked so far are thermodynamically preferred in any relevant region of the phase diagram predicted within those holographic models.

Our study also opens the door to calculating transport properties of nuclear matter. This can already be done with the well-known $D^0$ configuration, but it is to be expected that the calculation might be facilitated in the $D_{RL}$ phase due to its trivial infrared and ultraviolet behavior (and being non-singular at the same time). Strong-coupling results for instance for the neutrino emissivity would be of particular astrophysical interest. Such a study would first require the extension to nonzero isospin charge, while the calculation of, for instance, the shear viscosity of dense holographic nuclear matter can be done within the present isospin-symmetric approach. Further ideas concern anisotropic variants of the phases found here, for instance the chiral density wave, including a comparison to field-theoretic studies \cite{Papadopoulos:2024agt}, or the inclusion of a magnetic field, for which holographic studies within the pointlike approximation already exist \cite{Preis:2011sp}.

\acknowledgments

We thank Lorenzo Bartolini and Niko Jokela for useful comments and discussions as well as Domenico Logoteta and Albino Perego for sharing with us the nuclear matter data shown in figure \ref{fig:KnB}.
C.~E.~acknowledges support by the Deutsche Forschungsgemeinschaft (DFG, German Research Foundation) through the CRC-TR 211 'Strong-interaction matter under extreme conditions'-- project number 315477589 -- TRR 211.
The work of N.K. is supported by CONICET.

\appendix

\section{Pointlike approximation}
\label{app:pointlike}

In this appendix we discuss baryonic matter in the pointlike approximation with an arbitrary number of pointlike layers in the bulk. We employ an analogous notation as for the multi-jump finite-width calculation in the main text, i.e., we allow for a
 pointlike baryon layer at $u=u_{\rm KK}$ and for $d$ layers on one half of the flavor branes at locations $u_1<\ldots<u_d$. For this general case, the Lagrangian is 
\be \label{Lpoint}
{\cal L} = -\frac{u^{5/2}\sqrt{f}A'^2}{2} + \left[\bar{n}_c\delta(u-u_{\rm KK}) +\sum_{k=1}^d \bar{n}_k\delta(u-u_k)\right]\left[\frac{u}{3}-A(u)\right] \, .
\ee
Here, $\bar{n}_c$ and $\bar{n}_k$ are the baryon densities at $u=u_{\rm KK}$ and $u=u_k$, respectively\footnote{Since we are working on one half of the flavor branes, it might seem more sensible to replace $\bar{n}_c$ by $\bar{n}_c/2$. This redefinition is irrelevant for the physics, however, since $\bar{n}_c$ is determined dynamically. We thus keep the definition of eq.\ (\ref{Lpoint}) for notational simplicity.}. We shall allow for configurations where $\bar{n}_c=0$ (even number of points with baryon sources on both halves of the branes) and with nonzero $\bar{n}_c$ (odd number). In the same notation as in the main text, $A(u)$ is the temporal component of the abelian gauge field with boundary condition $A(\infty)=\bar{\mu}_B$, and the term proportional to $A'^2$ is the YM contribution. The terms proportional to $A(u)$ originate from the CS term, whereas the terms proportional to $u/3$ are mass terms, computed from the action of the baryonic D4-branes wrapped on the internal $S^4$. Such mass terms are not explicitly present in the finite-width approach of the main part of this paper, where the energy of the baryons is generated by the non-abelian gauge field in the YM contribution through the non-singular function $h(u)$. In our dimensionless units the Lagrangian (\ref{Lpoint}) does not depend on $\lambda$. Therefore, in the pointlike case the scaling with {\it both} parameters of the model, $\lambda$ and $M_{\rm KK}$, is done trivially after the numerical calculation; this is contrast to the main part, where the Lagrangian and the solutions to the equations of motion do depend on $\lambda$ (but not on $M_{\rm KK}$).   

\subsection{General solution}

The integrated equation of motion for $A(u)$ is 
\be \label{eomPoint}
A' = \frac{\bar{n}_c + \sum_{k=1}^d \bar{n}_k \Theta(u-u_k)}{u^{5/2}\sqrt{f}}
\, .
\ee
Although $A(u)$ is no longer smooth in the presence of pointlike sources, we shall still construct the solution such that $A(u)$ is continuous. 
The chemical potential assumes the form 
\be \label{muBPoint}
\bar{\mu}_B = A_d + \frac{\bar{n}_B}{3u_{\rm KK}^{3/2}}\left(\pi-2\arctan \tilde{z}_d\right) \, ,
\ee
where the total density is
\be \label{nBnc}
\bar{n}_B = \bar{n}_c + \sum_{k=1}^d \bar{n}_k \, ,
\ee
and where we have introduced the rescaled variable
\be
\tilde{z}\equiv \frac{z}{u_{\rm KK}} \, .
\ee
The variables of the action are the partial densities $\bar{n}_c,\bar{n}_1,\bar{n}_2,\ldots,\bar{n}_d$ and the locations $u_1,\ldots, u_d$, which all have to be determined by finding the stationary points of the effective potential
\be \label{OmPoint0}
\Omega  = -\int_{u_{\rm KK}}^\infty du\frac{u^{5/2}\sqrt{f}A'^2}{2}+\bar{n}_c\frac{u_{\rm KK}}{3}+\sum_{k=1}^d \bar{n}_k\frac{u_k}{3} - \bar{n}_c A_c - \sum_{k=1}^d \bar{n}_kA_k\, .
\ee
Before taking the derivatives with respect to the partial densities and locations of the sources, we compute the various terms explicitly by using the equation of motion (\ref{eomPoint}). 
Firstly, 
\be
\int_{u_{\rm KK}}^\infty du\frac{u^{5/2}\sqrt{f}A'^2}{2} = \frac{\bar{n}_B^2\pi}{6u_{\rm KK}^{3/2}}- \frac{1}{3u_{\rm KK}^{3/2}}\sum_{k=1}^d \bar{n}_k\left[\bar{n}_k+2\left(\bar{n}_c+\sum_{\ell=1}^{k-1} \bar{n}_\ell\right)\right]\arctan \tilde{z}_k \, .
\ee
Next, we need 
\bea \label{AkPoint}
A_k &=& \bar{\mu}_B - \int_{u_k}^\infty du\, A' \non[2ex]
&=& \bar{\mu}_B - \frac{\bar{n}_B\pi}{3u_{\rm KK}^{3/2}} + \frac{2}{3u_{\rm KK}^{3/2}}\left[\left(\bar{n}_B-\sum_{\ell=k}^d \bar{n}_\ell\right)\arctan \tilde{z}_k +\sum_{\ell=k}^d \bar{n}_\ell\arctan \tilde{z}_\ell \right] \, , 
\eea
and, making use of this result for $k=1$,  
\bea \label{AcPoint}
A_c &=& A_1 - \int_{u_{\rm KK}}^{u_1} du\, A' 
= \bar{\mu}_B- \frac{\bar{n}_B\pi}{3u_{\rm KK}^{3/2}} + \frac{2}{3u_{\rm KK}^{3/2}}\sum_{\ell=1}^d \bar{n}_\ell\arctan \tilde{z}_\ell  \, .
\eea
From eqs.\ (\ref{AkPoint}) and (\ref{AcPoint}) we obtain 
\bea
&&\bar{n}_c A_c + \sum_{k=1}^d \bar{n}_kA_k = \bar{\mu}_B \bar{n}_B - \frac{\bar{n}_B^2\pi}{3u_{\rm KK}^{3/2}} \non[2ex]
&& + \frac{2}{3u_{\rm KK}^{3/2}}\left[(\bar{n}_B+\bar{n}_c)\sum_{k=1}^d \bar{n}_k\arctan \tilde{z}_k +\sum_{k=1}^d \bar{n}_k\sum_{\ell=k}^d \bar{n}_\ell(\arctan \tilde{z}_\ell -\arctan \tilde{z}_k)\right] \, .
\eea
Consequently, the explicit form of the effective potential (\ref{OmPoint0}) is
\bea \label{OmPoint}
\Omega &=& -\bar{\mu}_B \bar{n}_B + \bar{n}_c\frac{u_{\rm KK}}{3}+\sum_{k=1}^d \bar{n}_k\frac{u_k}{3} + \frac{\bar{n}_B^2\pi}{6u_{\rm KK}^{3/2}}  \non[2ex]
&&- \frac{1}{3u_{\rm KK}^{3/2}}\sum_{k=1}^d \bar{n}_k\left[(2\bar{n}_c+\bar{n}_k)\arctan \tilde{z}_k+2\sum_{\ell=k+1}^d \bar{n}_\ell \arctan \tilde{z}_\ell\right] \, .
\eea
We can now easily compute stationarity with respect to $u_r$, $\bar{n}_c$, $\bar{n}_r$, respectively,  
\begin{subequations}
\bea
 0&=& 2\bar{n}_c+\bar{n}_r+2\sum_{k=1}^{r-1}\bar{n}_k -\frac{2u_{\rm KK}^{3/2}}{3}\tilde{z}_ru_r \, , \label{statur}\\[2ex] 
\bar{\mu}_B &=& \frac{u_{\rm KK}}{3} + \frac{\bar{n}_B\pi}{3u_{\rm KK}^{3/2}}- \frac{2}{3u_{\rm KK}^{3/2}}\sum_{k=1}^d \bar{n}_k\arctan \tilde{z}_k  \, , \label{statnc} \\[2ex] 
\bar{\mu}_B &=& \frac{u_r}{3} + \frac{\bar{n}_B\pi}{3u_{\rm KK}^{3/2}}- \frac{2}{3u_{\rm KK}^{3/2}}\left[\left(\bar{n}_c+\sum_{k=1}^r \bar{n}_k\right)\arctan \tilde{z}_r + \sum_{k=r+1}^d \bar{n}_k \arctan \tilde{z}_k \right] \, . \label{statnr}
\eea
\end{subequations}
Here, (\ref{statur}) and (\ref{statnr}) are  $d$ equations for $r=1,\ldots,d$, and eq.\ (\ref{statnc}) only exists as a condition if there is an odd number of baryon sources (i.e., if $\bar{n}_c\neq 0$). 

These equations can be brought into a convenient form as follows. First, we separate the $r=1$ case from eq.\ (\ref{statur}) and combine the remaining equations pairwise for $r$ and $r+1$, such that we can write 
\begin{subequations}\label{ncn0}
\bea
2\bar{n}_c+\bar{n}_1 &=& \frac{2u_{\rm KK}^{3/2}}{3}  \tilde{z}_1u_1 \, , \\[2ex]
\bar{n}_{r+1} + \bar{n}_r &=& \frac{2u_{\rm KK}^{3/2}}{3}(\tilde{z}_{r+1}u_{r+1}-\tilde{z}_ru_r) \, ,
\eea
\end{subequations}
where $r=1,\ldots, d-1$. Next, we combine the $r=1$ case from (\ref{statnr}) with (\ref{statnc}) and again combine the remaining equations pairwise for $r$ and $r+1$, 
\begin{subequations}\label{ncarc}
\bea
\bar{n}_c  &=& \frac{u_{\rm KK}^{3/2}}{2}\frac{u_1-u_{\rm KK}}{\arctan \tilde{z}_1} \, , \label{nc} \\[2ex]
\bar{n}_c+\sum_{k=1}^r \bar{n}_k&=& \frac{u_{\rm KK}^{3/2}}{2}\frac{u_{r+1}-u_r}{\arctan \tilde{z}_{r+1}-\arctan \tilde{z}_r}   \, ,
\eea
\end{subequations}
where $r=1,\ldots, d-1$.
We now use eqs.\ (\ref{ncn0}) and (\ref{ncarc}) to derive expressions for the individual densities and relations between the locations $u_r$ where the densities are eliminated,
\begin{subequations}
\bea \label{nr3}
\bar{n}_r &=& \frac{2u_{\rm KK}^{3/2}}{3}\left[\tilde{z}_ru_r+2\sum_{k=1}^{r-1}(-1)^{r+k} \tilde{z}_k u_k\right] + 2(-1)^r\bar{n}_c  \, , \label{nruKK} \\[2ex]
\frac{u_{r+1}-u_r}{\arctan \tilde{z}_{r+1}-\arctan \tilde{z}_r} &=& \frac{4}{3}\sum_{k=1}^r (-1)^{r+k} \tilde{z}_ku_k + 2(-1)^r\frac{\bar{n}_c}{u_{\rm KK}^{3/2}} \, , \label{uarctan}
\eea
\end{subequations}
where $r=1,\ldots,d$ in eq.\ (\ref{nruKK}) and $r=1,\ldots,d-1$ in eq.\ (\ref{uarctan}). Without baryons at the tip $\bar{n}_c=0$, otherwise   $\bar{n}_c$ is given by eq.\ (\ref{nc}). 
For given $d$ and $u_1$, the set of equations (\ref{uarctan}) yields all $u_2, \ldots, u_d$. These equations have to be solved numerically, but they decouple and can be solved successively from $u_1$ up to $u_d$. The results can then be inserted into (\ref{nruKK}) to compute all densities $\bar{n}_1,\ldots, \bar{n}_d$, and thus also the total density $\bar{n}_B$ via eq.\ (\ref{nBnc}). 
The chemical potential is then obtained from setting $r=d$ in eq.\ (\ref{statnr}), which results in the simple form  (\ref{muBPoint}) with $A_d=u_d/3$. All results can then be inserted into eq.\ (\ref{OmPoint}) to compute the free energy density. Repeating this 
procedure for many $u_1$ yields the full branch of solutions for the given number of pointlike baryon layers. 

We refer to the phase with $\ell$ layers as $P_\ell$, where the total number of layers is $\ell=2d$ or $\ell=2d+1$ depending on whether there is a layer at the tip of the flavor branes. Results for $P_1$ and  $P_2$ are used in the main text in figures \ref{fig:width} -- \ref{fig:KnB}. Our equations allow us to evaluate any number of layers with very little numerical effort, and the results for a large number of layers lead us to the following discussion of the limit of infinitely many layers.     

\subsection{Continuum limit}
\label{sec:continuum}

We find numerically that the solution converges as the number of layers goes to infinity. More precisely, for a fixed $\bar{\mu}_B$, we find that the distance between the locations of the delta-peaks $u_1,\ldots,u_d$ goes to zero with $u_1\to u_{\rm KK}$, and the location of the ``last'' peak $u_d$ goes to a finite value that depends on $\bar{\mu}_B$. Let us denote this value by $u_\infty$ and, correspondingly, in the $\tilde{z}$ coordinate, by $\tilde{z}_\infty$. As a consequence, the thermodynamic quantities $\bar{n}_B$ and $\Omega$ also asymptote to finite values. In fact, we can derive the infinite-layer result by taking the continuum limit of the finite-layer result. To this end, we first define the discrete charge distribution
\be \label{rhozk}
\rho_k\equiv \frac{1}{2\bar{n}_B}\frac{\bar{n}_k}{\Delta \tilde{z}_k} \, , \qquad 
\Delta \tilde{z}_k \equiv \tilde{z}_k-\tilde{z}_{k-1} \, , 
\ee
with continuum limit 
\be
\rho(\tilde{z}) = \lim_{\Delta z_k\to 0} \rho_k  \, .
\ee
As in the main part of the paper, we have normalized the charge distribution by the total dimensionless charge (recall that $\bar{n}_B$ should be understood as the dimensionless charge on one half of the flavor branes), such that 
\be \label{intrhoinf}
 2\int_{0}^{\tilde{z}_\infty} d\tilde{z} \, \rho(\tilde{z}) = 1 \, .
\ee
To calculate the continuum limit, we first insert eq.\ (\ref{uarctan}) into eq.\ (\ref{nruKK}),
\bea \label{nr2}
\bar{n}_r &=& \frac{2u_{\rm KK}^{3/2}}{3}\left[-\tilde{z}_ru_r+\frac{3}{2}\frac{u_{r+1}-u_r}{\arctan \tilde{z}_{r+1}-\arctan \tilde{z}_r}\right] \, .
\eea
Next, we abbreviate 
\be
p(u) \equiv \arctan\tilde{z}(u) \, , 
\ee
and write the Taylor expansion of this function in the form
\be
\frac{u_{r+1}-u_r}{p(u_{r+1})- p(u_r)} = \frac{1}{p'}  - \frac{u_{r+1}-u_r}{2}\frac{p''}{p'^2} + {\cal O}[(u_{r+1}-u_r)^2] \, . 
\ee
Inserting this into eq.\ (\ref{nr2}) and using the definition (\ref{rhozk}) yields the continuum charge distribution, 
\bea
\rho(\tilde{z}) = -\frac{u_{\rm KK}^{3/2}}{4\bar{n}_B}\frac{\partial u}{\partial \tilde{z}} \frac{p''}{p'^2} = \frac{u_{\rm KK}^{5/2}}{18\bar{n}_B}\frac{3+5\tilde{z}^2}{(1+\tilde{z}^2)^{2/3}} \, .
\eea
We can easily integrate over this function to obtain a relation between $\bar{n}_B$ and $\tilde{z}_\infty$ via eq.\ (\ref{intrhoinf}),
\be
\bar{n}_B = \frac{u_{\rm KK}^{3/2}\tilde{z}_\infty u_\infty}{3} \, , 
\ee
which, using eq.\ (\ref{muBPoint}), yields the chemical potential  
\be
\bar{\mu}_B = \frac{u_\infty}{3}\left[1+\frac{\tilde{z}_\infty}{3}(\pi-2\arctan\tilde{z}_\infty)\right] \, . 
\ee
For a given $\bar{\mu}_B$, this relation can be used to determine $\tilde{z}_\infty$ numerically. Finally, inserting all these results into eq.\ (\ref{OmPoint}), we compute the continuum limit of the free energy as a function solely of $\tilde{z}_\infty$, 
\allowdisplaybreaks
\bea
\Omega &=& -\bar{\mu}_B\bar{n}_B +\frac{1}{3}\int_0^{\tilde{z}_\infty} d\tilde{z} \,\rho(\tilde{z})u(\tilde{z}) + \frac{\pi\bar{n}_B^2}{6u_{\rm KK}^{3/2}} - \frac{2}{3u_{\rm KK}^{3/2}} \int_0^{\tilde{z}_\infty} d\tilde{z} \,\rho(\tilde{z})\int_{\tilde{z}}^{\tilde{z}_\infty}dv\,\rho(v)\arctan v \non[2ex]
&=& -\frac{u_{\rm KK}^{7/2}}{63}\tilde{z}_\infty \left\{\frac{u_\infty^2}{u_{\rm KK}^2}+\frac{7\tilde{z}_\infty u_\infty^2}{6u_{\rm KK}^2}\left(\pi-2\arctan\tilde{z}_\infty\right)-{}_2F_1\left[\frac{1}{3},\frac{1}{2},\frac{3}{2},-\tilde{z}_\infty^2\right]\right\} \, ,
\eea
where ${}_2F_1$ is the hypergeometric function. 

\begin{figure} [t]
\begin{center}
\hbox{\includegraphics[width=0.5\textwidth]{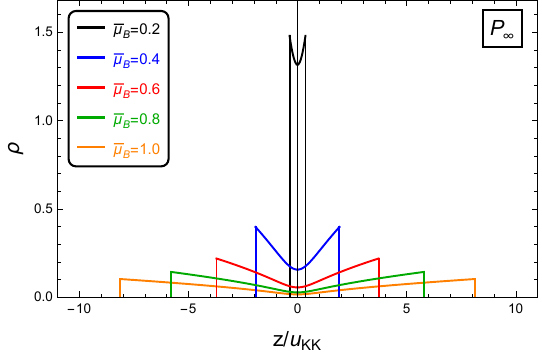}\includegraphics[width=0.5\textwidth]{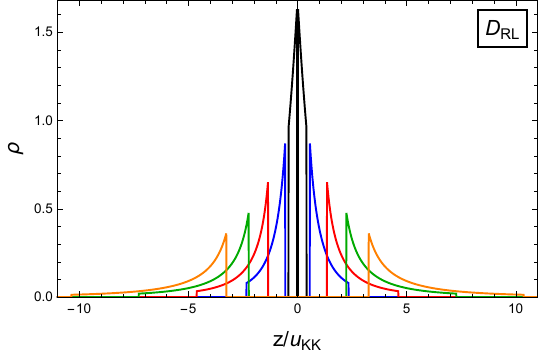}}  
\caption{{\it Left panel:} Charge distribution of the $P_\infty$ phase -- the continuum limit of the multi-layer pointlike solution -- for 5 different values of the dimensionless chemical potential $\bar{\mu}_B$. {\it Right panel:} As a comparison, charge distribution of the $D_{RL}$ phase from the main text for $\lambda=7.09$, for the same chemical potentials.}  
\label{fig:rhoInf}
\end{center}
\end{figure}

These results are used to compute the infinite-layer results in figures \ref{fig:omega2} and \ref{fig:KnB} in the main text. 
To illustrate the behavior of this phase -- which we term $P_\infty$ -- we show the charge distribution  for different chemical potentials in figure \ref{fig:rhoInf} (left panel), in comparison to the $D_{RL}$ phase (right panel). One can check that for a finite number of layers $\ell$ the discrete charge distribution, defined via eq.\ (\ref{rhozk}), follows these curves approximately and approaches them as $\ell\to\infty$.

\bibliographystyle{JHEP}
\bibliography{refs}

\end{document}